\documentclass[JHEP,12pt]{article}

\usepackage{subfigure}
\usepackage{amssymb,amsmath}
\usepackage{braket}
\usepackage{jheppub}
\usepackage{mathrsfs}
\usepackage{amsthm}
\usepackage{bbold}
\usepackage{comment} 
\author[1]{Raphael Bousso}
\author[2]{and Arvin Shahbazi-Moghaddam}

\affiliation[1]{Berkeley Center for Theoretical Physics and Department of Physics,\\
University of California, Berkeley, CA 94720, USA} 

\affiliation[2]{Stanford Institute for Theoretical Physics,\\ Stanford University, Stanford, CA 94305, USA}

\emailAdd{bousso@berkeley.edu}
\emailAdd{arvinshm@gmail.com}

\title{Island Finder and Entropy Bound}
\abstract{Identifying an entanglement island requires exquisite control over the entropy of quantum fields, which is available only in toy models. Here we present a set of sufficient conditions that guarantee the existence of an island and place an upper bound on the entropy computed by the island rule. This is enough to derive the main features of the Page curve for an evaporating black hole in any spacetime dimension. Our argument makes use of Wall's maximin formulation and the Quantum Focusing Conjecture. As a corollary, we derive a novel entropy bound.}

\begin{document}
\maketitle

\section{Introduction}

\subsection{Entanglement Wedges and Islands}

The quantum-corrected~\cite{Engelhardt:2014gca}, covariant~\cite{Hubeny:2007xt} Ryu-Takayanagi~\cite{Ryu:2006bv} (RT) prescription computes the CFT entropy of a boundary region in terms of a dual asymptotically AdS bulk spacetime. Originally an {\em ad hoc} proposal, it follows under certain assumptions from a Euclidean gravitational path integral~\cite{Lewkowycz:2013nqa, Faulkner:2013ana, Dong:2017xht, Penington:2019kki, Almheiri:2019qdq}. This derivation implies that the RT prescription is not tied to the AdS/CFT correspondence but can be evaluated in any spacetime $M$.

Indeed, RT yields the Page curve~\cite{Page:1993wv} for the entropy of the {\em bulk} radiation emitted by a black hole~\cite{Penington:2019npb, Almheiri:2019psf}. The bulk state and geometry are treated semiclassically. In this approximation, the radiation is thermal~\cite{Hawking:1974sw}, and its von Neumann entropy $S(R)$ increases monotonically, implying information loss~\cite{Hawking:1976ra}. Using the same semiclassical solution, the RT proposal computes the radiation entropy differently, as the generalized entropy\footnote{For a partial Cauchy surface $X\subset M$, $S_{\rm gen}(X) = \text{Area}[\partial X]/4G\hbar + S(X)$, where $\partial$ denotes the boundary of a set, and $S(X)$ is the renormalized von Neumann entropy of the density operator of the quantum field theory state reduced to $X$. See the appendix in Ref.~\cite{Bousso:2015mna} for a detailed discussion of this quantity.} of the {\em entanglement wedge} of the radiation, $E(R)$:
\begin{equation}
    S(\mathbf{R}) = S_{\rm gen}[E(R)]~.
    \label{srer}
\end{equation}
The bold-face notation~\cite{Almheiri:2019yqk} distinguishes the (presumably correct) entropy computed by RT from the entropy $S(R)$ computed directly from the semiclassical radiation state (See \cite{Bousso:2019ykv, Bousso:2020kmy} for a proposal to reconcile the bold and unbold states.)

The original RT prescription defines an entanglement wedge for regions on the conformal boundary of AdS. In the present context, $R$ is a bulk system, and the entanglement wedge must be defined as follows~\cite{Bousso:2020kmy}:
\begin{enumerate}
    \item $E(R)=I\cup R$, where $I\subset M$; 
    \item $S_{\rm gen}(I\cup R)$ is stationary under any local variations of the boundary surface $\partial I$; 
    \item among all such regions globally, $I$ yields the smallest $S_{\rm gen}(I\cup R)$.
\end{enumerate}

The above definitions apply if $R$ is a nongravitating system external to $M$; in asymptotically AdS geometries the radiation can be extracted into such a system~\cite{Penington:2019npb, Almheiri:2019psf}. We now turn to the case where $R$ is a weakly gravitating region inside the spacetime. For example, $R$ may be a distant region occupied by Hawking radiation in an asymptotically flat or AdS spacetime (see Ref.~\cite{Bousso:2019ykv} for a detailed setup).

Physically, one expects the RT prescription for a weakly gravitating region to reduce to that for a nongravitating system, Eq.~\eqref{srer}, and we shall assume this here. $R$ resides in a weakly gravitating region far from any potential island $I$, so $\partial I \cap \bar R=\varnothing$, where an overbar denotes topological closure. As before, stationarity of $S_{\rm gen}$ is required only under variations of $\partial I$, not of $\partial R$. (This can be implemented in a path integral derivation~\cite{Dong:2020uxp}.) Thus, the definition of $E(R)$ is essentially unchanged.

However, in the presence of gravity, it is simplest to work with the generalized entropy of the region $\mathbf{R}$ (a cutoff-independent quantity), so we add its boundary area to both sides of Eq.~\eqref{srer}:
\begin{equation}
    S_{\rm gen}(\mathbf{R}) = S_{\rm gen}[E(R)]~.
\end{equation} 

It is easy to derive the Page curve, at least approximately, {\em if one ignores condition 2}. We will now summarize this incomplete argument, before discussing how it can be completed.

\begin{figure}[t]
\begin{center}
  \includegraphics[scale=.7]{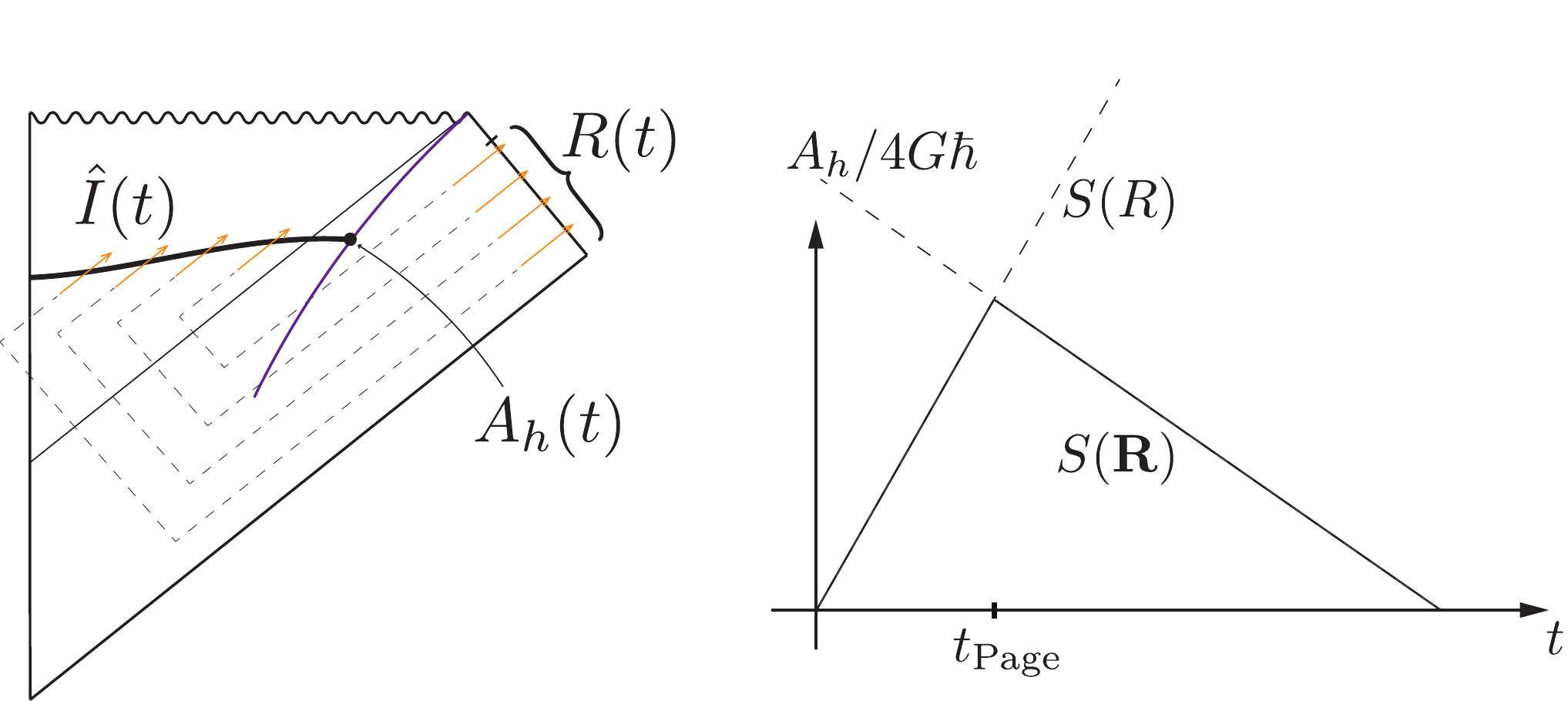} 
\end{center}
\caption{Left: evaporating black hole; right: its Page curve. After the Page time, the semiclassical entropy $S(R)$ of the Hawking radiation in the asymptotic region $R$ exceeds the Bekenstein-Hawking entropy of the black hole, $A_h/4G\hbar$. The ``Hawking partners'' in the black hole interior purify $R$. (Dashed lines indicate entanglement.) Therefore, adjoining $\hat I$ to $R$ decreases the generalized entropy $S_{\rm gen}$. However, islands must have stationary $S_{\rm gen}$. Solving for this condition exceeds present analytic control over the entropy. The island finder presented here sidesteps this obstruction.}
\label{fig:one-sided2}
\end{figure}

The Page time $t_{\rm Page}$ is defined as the time when the black hole and radiation entropies are equal in the semiclassical analysis:
\begin{equation}
    S[R(t_{\rm Page})]=\frac{A_h(t_{\rm Page})}{4G\hbar}~.\label{tpage}
\end{equation}
Let $\hat I(t)$ be the black hole interior at time $t$ (see Fig.~\ref{fig:one-sided2}). The Hawking ``partners'' in $\hat I(t)$ purify the radiation $R(t)$ emitted so far; hence
\begin{equation}
    S_{\rm gen}[\hat I(t)\cup R]\approx \frac{A_h(t)}{4G\hbar}~.\label{bharea}
\end{equation} 
Before the Page time, this is greater than $S(R)$ by definition, so $\hat I$ is not a viable island candidate; one finds that $I(t)=\varnothing$, $E(R)=R$, and $S(\mathbf{R})=S(R)$. This corresponds to the rising part of the Page curve, where it agrees with Hawking's curve. But after the Page time, $S_{\rm gen}[\hat I(t)\cup R]<S(R)$ by Eqs.~\eqref{tpage} and \eqref{bharea}. Thus, the inclusion of an island $I(t)\approx \hat I(t)\neq\varnothing$ is favored, and we have $S(\mathbf{R})=S(I\cup R)\approx A_h/4G\hbar$. As the black hole evaporates and its horizon shinks, this yields the decaying part of the Page curve required by unitarity.

The Page curve result has been extended to asymptotically flat spacetimes~\cite{Gautason:2020tmk, Hartman:2020swn}, settings with two layers of holography~\cite{Almheiri:2019hni, Almheiri:2019psy}, and eternal black holes~\cite{Almheiri:2019yqk}. Entanglement islands can also appear in cosmology, where their significance is less obvious~\cite{Chen:2020tes, Hartman:2020khs}.

\subsection{Summary and Outline}

Our brief summary of the Page curve result has a major gap. We explained why condition 3 (global minimization of $S_{\rm gen}$) favors inclusion of the black hole interior $\hat I(t)$ in $E(R)$ after the Page time. However, we did not show that condition 2 (local extremality) can be satisfied by some deformation of $\hat I(t)$ small enough to preserve condition 3.

One way to fill this gap is to find $I(t)$ exactly, and to verify condition 2. However, explicit solutions of the quantum extremality condition have been found only in $1+1$ bulk dimensions~\cite{Almheiri:2019psf, Almheiri:2019hni, Almheiri:2019yqk}, or in toy models of higher-dimensional black holes~\cite{Penington:2019npb}. The difficulty lies in computing the von Neumann entropy $S(I\cup R)$. This depends on the detailed state of the dynamics of the radiation fields, including modes with nonzero angular momentum, and their interactions. Even free fields scatter nontrivially in a black hole background, placing an exact calculation out of reach.

In Sec.~\ref{sci}, we develop an alternative way to ensure that condition 2 holds. We show that the {\em existence} of a suitable island $I$ follows from simple sufficient conditions that are easy to verify:\footnote{Ref.~\cite{Penington:2019npb} presents an elegant existence argument that establishes an island in the setting of an evaporating black hole. It makes use of properties of the event horizon and is inequivalent to the more general argument presented here.} {\em Let $I'$ be a region that (i) satisfies condition 1, $S_{\rm gen}(I'\cup R)<S(R)$, and suppose that (ii) the generalized entropy of $I'\cup R$ increases under any small outward deformation of $I$, or decreases under any such deformation. Then there exists an island, $I\neq\varnothing$; and moreover}
\begin{equation}
    S(\mathbf{R})=S_{\rm gen}(I\cup R)\leq S_{\rm gen}(I'\cup R)~. 
\label{boundsr}
\end{equation} 

We will illustrate in a number of examples that finding a suitable $I'$ is not difficult; in particular, it suffices to understand the scaling of corrections to simple models of the entropy. Moreover, the upper bound \eqref{boundsr} is powerful enough to establish the main features of the Page curve for an evaporating black hole.

In Sec.~\ref{neb}, we consider a different but related problem that yields to a similar analysis. We consider a spacetime and (internal or external) reference system $R$ in a pure state. We show that the true entropy $S(\mathbf{R})$ cannot exceed the generalized entropy of appropriate bulk regions. For example, if $R$ is external, and $M$ is an evaporating black hole spacetime, an upper bound on $S(\mathbf{R})$ is furnished by the generalized entropy of the bulk region that can be probed by an asymptotic observer (the black hole exterior).

\section{Sufficient Conditions for Islands}
\label{sci}

In this section we identify sufficient conditions for the existence of an island.  In Sec.~\ref{ext} we begin with the case of an external reference system, $R\cap M=\varnothing$. In Sec.~\ref{dist} we allow $R$ to intersect with $M$. In Sec.~\ref{examples} we consider various examples in which our conditions easily establish the presence of islands; in particular, we show that they suffice to derive the Page curve.

\subsection{External Reference System} \label{ext}

Let $M$ be a semiclassical spacetime which together with a reference system $R$ is in a pure state. We take $R$ to be external to $M$. For definiteness, we assume that $R$ is nongravitating; otherwise, simply substitute $S\to S_{\rm gen}$ when the argument contains $R$.
\begin{figure}
\centering
\begin{subfigure}
  \centering
  \includegraphics[width=.58\linewidth]{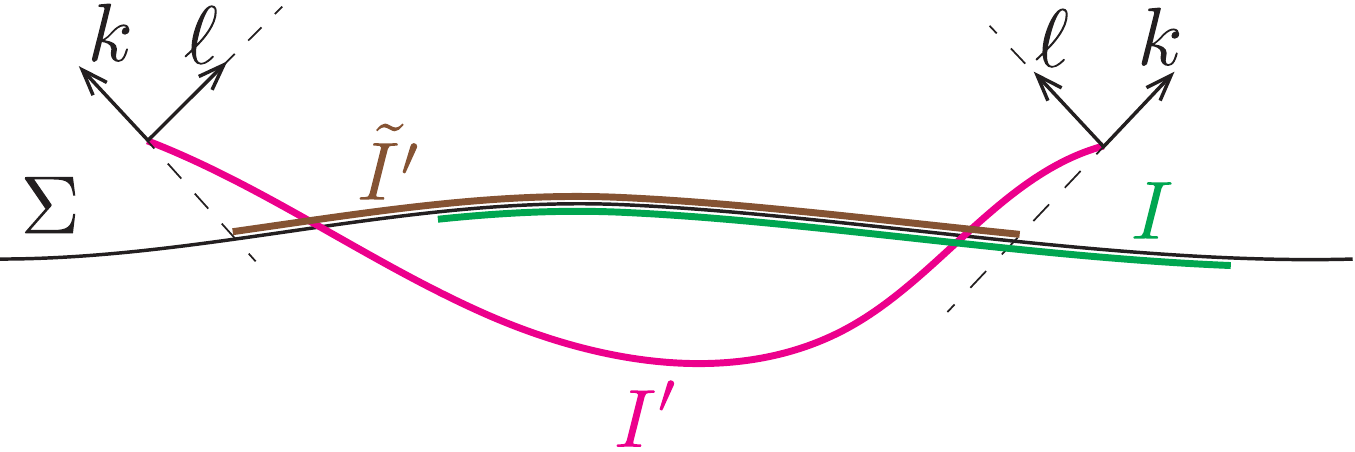}
  \label{fig:maximin1}
\end{subfigure}%
\begin{subfigure}
  \centering
  \includegraphics[width=.58\linewidth]{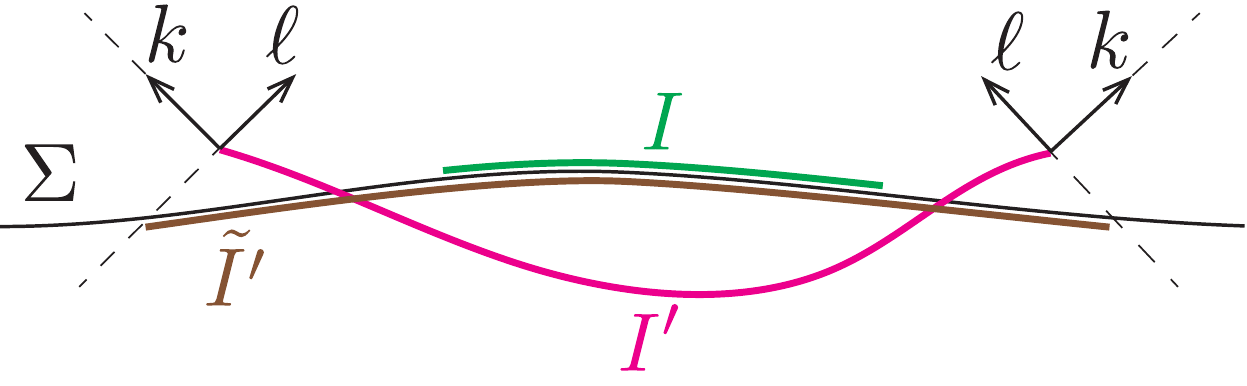}
  \label{fig:maximin2}
\end{subfigure}
\caption{Island finder. Suppose that $I'\cup R$ is quantum normal (top) or anti-normal (bottom). Then the generalized entropy of $I'\cup R$ decreases along the dashed lines to $\tilde I'\subset \Sigma$. An island $I$ with even smaller generalized entropy $S_{\rm gen}(I\cup R)$ must exist on the maximin Cauchy slice $\Sigma$. If $S_{\rm gen}(I'\cup R)<S(R)$, the island cannot be empty.}
\label{fig:test}
\end{figure}

Suppose there exists a partial Cauchy surface $I'\subset M$ satisfying the following conditions:
\begin{align}
    (i) & ~~~S_{\text{gen}}(I'\cup R)  < S(R)~; \label{reduce}\\
    (ii) & \begin{cases}
    k^{\mu} \Theta_{\mu}[I'\cup R] \geq 0~,~~ \ell^{\mu} \Theta_{\mu}[I'\cup R]  \leq 0~; \label{normal}\\
    \text{~~~~~~~~~~~~~~~~~~~~~~or}\\
    k^{\mu} \Theta_{\mu}[I'\cup R] \leq 0~,~~ \ell^{\mu} \Theta_{\mu}[I'\cup R]  \geq 0 ~,
   \end{cases}
\end{align}
where $\Theta$ is the quantum expansion one-form~\cite{Engelhardt:2014gca,Bousso:2015mna} and $k$ and $\ell$ are the outward and inward future-directed null vectors normal to $\partial I'$. Thus, for example, $k^\mu\Theta_\mu[I'\cup R, y]$ is the rate of change of $S_{\rm gen}$, per unit transverse area and unit affine parameter length, as $I'\cup R$ is deformed outward along the future-directed null geodesic orthogonal to $I'$ at $y$. We drop the argument $y$ when an equation holds for all $y$.

The first condition states that adjoining $I'$ to $R$ decreases the entropy, even as the Bekenstein-Hawking entropy of $I'$ is included. The second, Eq.~\eqref{normal}, says that $I'\cup R$ is {\em quantum normal} (its generalized entropy does not decrease under any small outward deformation of $I'$) or {\em quantum anti-normal} (the opposite).

We will now show that these conditions imply the existence of a non-empty island region $I$. Our proof uses the maximin construction of the HRT surface~\cite{Wall:2012uf}, which we assume extends to a quantum maximin prescription~\cite{Akers:2019lzs}: on every Cauchy surface of $M$, one finds the region $I''$ that minimizes $S_{\rm gen}(I''\cup R)$. (Note that $I''$ may be the empty set.) One then maximizes the same quantity over all Cauchy surfaces of $M$. The island $I$ is defined to be the region $I''$ that achieves this maximum. (This is expected to exist~\cite{Wall:2012uf,Akers:2019lzs}.)

Note that we define $I$ as an achronal region; hence it is non-unique. Similarly, the maximin Cauchy slice $\Sigma$ is non-unique. The relevant unique object is the domain of dependence $D(I)$. (We ignore the degenerate case where there are two islands with identical generalized entropy but different $D$.) Any other Cauchy slice of $D(I)$ will be an island if $I$ is, though not every Cauchy slice of $D(I)$ will be part of a maximin slice $\Sigma$.  

Our goal is to show that $I\neq \varnothing$. In the normal case, $k^{\mu} \Theta_{\mu}[I'\cup R]\geq 0$, we define 
\begin{equation}
    \tilde I' \equiv D(I') \cap \Sigma~,
\end{equation}
as the {\em representative} of $I'$ on $\Sigma$. In the anti-normal case, $k^{\mu} \Theta_{\mu}[I'\cup R]\leq 0$, we define the representative instead as
\begin{equation}
    \tilde I' \equiv J(I') \cap \Sigma~,
\end{equation}
where $J$ denotes all points that can be reached from $I'$ along a causal curve (the future and past of $I'$). In either case, note that $\tilde I'$ is obtained from $I'$ by deforming along an orthogonal null congruence with initially negative quantum expansion. We assume the Quantum Focussing Conjecture (QFC)~\cite{Bousso:2015mna}, that the quantum expansion cannot increase along a null shape deformation. This implies that 
\begin{equation}\label{eq-tildeI'I'}
    S_{\rm gen}(\tilde I'\cup R) \leq S_{\rm gen}(I'\cup R)~.
\end{equation}
Since $\Sigma$ is the maximin Cauchy slice,
\begin{equation}
    S_{\rm gen}(\tilde I'\cup R)\geq S_{\rm gen}(I\cup R)~.
\end{equation}
Combined with Eq. \eqref{eq-tildeI'I'}, this implies\footnote{This intermediate result is closely related to corollary 16b of \cite{Wall:2012uf}. A simple application of our argument to asymptotically AdS spacetimes yields the following result: given a partial Cauchy slice $A$ on the asymptotic boundary of AdS, let $X$ be the RT surface associated to $A$ with homology slice $H$. Now, consider another surface $X'$, homologous to $A$ with homology slice $H'$. If $H'$ is a quantum normal or anti-normal region, then $S_{\rm gen}(H)\leq S_{\rm gen}(H')$.}
\begin{equation}\label{eq-intermed31}
    S_{\rm gen}(I'\cup R)\geq S_{\rm gen}(I\cup R)~.
\end{equation}
Using the assumption in Eq.~\eqref{reduce}, we get
\begin{equation}
    S_{\rm gen}(I\cup R) < S_{\rm gen}(R)~.
\end{equation}
Therefore, we conclude that $I\neq \varnothing$.

\subsection{Distant Reference System} \label{dist}

In Section \ref{ext}, the reference system $R$ was external to the spacetime $M$. We will now allow a system that is wholly or partially inside the spacetime: $R\cap M\neq \varnothing$. 

In order to generalize our island finder to this setting, we shall require that there exists a partial Cauchy slice $I_0$ spacelike to $R$ such that $I_0 \cup R$ is a quantum normal region with respect to deformations at $\partial I_0$. That is, we require
\begin{align}
    \bar{I_0} & \subset M-\overline{J(R)}~,\\
    k^\mu \Theta_\mu[I_0 \cup R] & \geq 0~, \label{knormal} \\
    \ell^\mu \Theta_\mu[I_0 \cup R] & \leq 0~\label{lnormal}.
\end{align}
where as before an overbar denotes closure. As before, $\Theta_\mu[I_0 \cup R]$ is the quantum expansion one-form, and $k^\mu$ and $\ell^\mu$ are future-directed null vectors fields in the normal bundle of $\partial I_0$ in the outward and inward directions respectively.

For example, these conditions will be satisfied when $I_0$ is the interior of a sufficiently large, approximately round sphere in an asymptotically flat spacetime; and $R$ is the exterior of a slightly larger sphere concentric with the first, or any subsystem thereof (such as the Hawking radiation emitted by a black hole). Note that we do not require that $R$ be weakly gravitating, but in many examples of interest this will be the case. Also, we do not require that Eqs.~\eqref{knormal} and \eqref{lnormal} hold at $\partial R$.

Now, suppose there exists a partial Cauchy slice $I' \subset D(I_0)$ satisfying the conditions \eqref{reduce} and \eqref{normal}. That is, $I'\cup R$ is quantum normal or anti-normal, and adjoining $I'$ to $R$ reduces the generalized entropy of $R$. Then there exists a non-empty quantum extremal region $\hat I\subset D(I_0)$ satisfying $S_{\rm gen}(\hat{I} \cup R) < S_{\rm gen}(R)$. 

Note that this conclusion is weaker than in Sec.~\ref{ext}: $\hat I$ need not globally minimize $S_{\rm gen}(I\cup R)$ among all eligible regions, since we are restricting our search to a subset of $M$. However, the true entanglement wedge can only have lower entropy, so $S_{\rm gen}(\hat I\cup R)$ provides an upper bound. Note also that when $I' \cup R$ is quantum normal, we can set $I_0=I'$, so there is no need to identify a larger $I_0$ region.

\paragraph{Proof, part I} The proof strategy is similar to that in Section \ref{ext}, except that we wish to restrict our search to the closed set $D(I_0)$.\footnote{A maximin procedure restricted to entanglement wedges of AdS without reflecting boundary conditions was considered in~\cite{Marolf:2019bgj}. See also Appendix B of~\cite{Brown:2019rox} for a related discussion.} We will first assume strict version of conditions \eqref{knormal} and \eqref{lnormal}:
\begin{align}
    k^\mu \Theta_\mu[I_0 \cup R] & > 0~, \label{knormalstrict} \\
    \ell^\mu \Theta_\mu[I_0 \cup R] & < 0~\label{lnormalstrict},
\end{align}
Later, we will demonstrate how to relax this assumption back to conditions \eqref{knormal} and \eqref{lnormal}.

Let $\hat{I}$ be the maximin region of $D(I_0)\cup R$, and let $\Sigma$ be a Cauchy surface of $D(I_0)$ on which $\hat I$ minimizes the generalized entropy of $\hat I\cup R$ among all subregions of $\Sigma$. Without loss of generality, we may take $\hat I\subset \Sigma$: since any Cauchy surface of $D(\hat I)$ is an equally good maximin region, we set $\hat I\to \Sigma \cap D(\hat I)$. As in Ref.~\cite{Wall:2012uf}, we assume that $\hat I$ is {\em stable}: any nearby Cauchy surface $\Sigma'$ obtained by infinitesimal deformation of $\Sigma$ contains a locally minimal region $\hat I'$ infinitesimally close to $\hat I$ with $S_{\rm gen}(\hat I'\cup R)\leq S_{\rm gen}(\hat I\cup R)$.

It immediately follows from the analysis of Sec.~\ref{ext} that
\begin{align}
   S_{\rm gen}(\hat{I} \cup R) < S_{\rm gen}(R)~.
\end{align}
We will now show that $\partial \hat{I} \cap \partial D(I_0) = \varnothing$. This precludes the (unwanted) possibility that $I\cup R$ is maximin but not locally stationary. It follows that $\hat{I} \cup R$ is a quantum extremal region \cite{Wall:2012uf}.

$\partial D(I_0)$ is the disjoint union of three sets: $\partial I_0$, and two null hypersurfaces $N_{+\ell}$ and $N_{-k}$ that lie in the future and past of $I_0$ respectively. The latter sets are generated by future and past-directed null geodesics orthogonal to $\partial I_0$ in the inward direction which end at caustics or self-intersections \cite{Akers:2017nrr}. Let $\ell^\mu$ ($k^\mu$) be the normal vector field to $N_{+\ell}$ ($N_{-k}$)  obtained by parallel propagation of $\ell^\mu|_{\partial I_0}$ ($k^\mu|_{\partial I_0}$).

Let $\Sigma_M$ be a Cauchy surface of $M$ that intersects each null generator of $N_{+\ell}$ and $N_{-k}$ at most at one point. Let $\Sigma_N\equiv \Sigma_M\cap D(I_0)$. By Eqs.~\eqref{knormalstrict}, \eqref{lnormalstrict}, and the QFC~\cite{Bousso:2015mna},
\begin{align}
  k^\mu \Theta_\mu[\Sigma_N \cup R;p] & > 0
 ~~~\text{ for all } p\in N_{-k}\cap \partial I_0 ~;\label{knormal2} \\
  \ell^\mu \Theta_\mu[\Sigma_N \cup R;p] & < 0
  ~~~\text{ for all } p\in N_{+\ell}\cap \partial I_0~, \label{lnormal2}
\end{align}

Suppose for contradiction that there exists a point $q\in \partial D(I_0)\cap \partial \hat I$. The generator of $\partial D(I_0)$ that contains $q$, and hence its tangent vector $\ell^\mu$ or $k^\mu$, will be normal to $\partial \hat{I}$ at $q$. (If $q\in \partial I_0$, this statement holds for either generator emanating from $q$.) Since $N_{+\ell}$  ($N_{-k}$) is nowhere to the past (future) of $\hat{I}$, theorem 1 in Ref.~\cite{C:2013uza} implies
\begin{align}
  k^\mu \Theta_\mu[\hat I \cup R;q] & > 0
                         ~~~\text{ for } q\in N_{-k}\cap \partial I_0 ~;\label{knormal3} \\
  \ell^\mu \Theta_\mu[\hat I \cup R;q] & < 0
                         ~~~\text{ for } q\in N_{+\ell}\cap \partial I_0~. \label{lnormal3}
\end{align}

Suppose first that $q\in \partial I_0$. In this case the above expansions imply that a small inward deformation of $\hat I$ will decrease the generalized entropy, in contradiction with the minimality of $S_{\rm gen}(\hat I\cup R)$ among all subregions of the maximin Cauchy surface $\Sigma$. 

We will now demonstrate this rigorously, using the notion of a surface-orthogonal exponential map~\cite{Akers:2017nrr},
\begin{equation}
    \exp_K~:~NK \to M~, ~~ (p,v) \to c_{p,v}(1)~,
\end{equation}
which takes a vector $v$ in the normal bundle $NK$ of a smooth submanifold $K$ at the point $p$ to the point at affine distance 1 along the unique geodesic through $p$ with tangent vector $v$. We will use $\widetilde{\exp}$ to denote an exponential map in which the submanifold $\Sigma$ plays the role of $M$ in the above definition.

Let $\tilde v^\mu(p)$ be a smooth inward-directed vector field in the normal bundle to $\partial \hat I$, viewed as a submanifold of $\Sigma$. We define the continuous one-parameter family of inward deformations of $\hat{I}$ as the regions
\begin{align}
    \hat{I}(\epsilon) = \text{int}\{\widetilde{\exp}_{\partial \hat I}(p, \epsilon \tilde v^\mu(p)): p \in \partial \hat{I} \}~.
\end{align}
Similarly, in the manifold $M$ we define
\begin{align}
    i(\epsilon) = \hat I \cap \text{Int}\{\exp_{\partial \hat I}(p, \epsilon v^\mu(p)): p \in \partial \hat{I} \}~,
\end{align}
where $v$ is the push-forward of $\tilde v$ under the embedding map of $\Sigma$ into $M$, and $\text{Int}\,X$ denotes the spacetime region spacelike and interior to $X$. For sufficiently small $\epsilon$, both families are well-defined. Moreover,
\begin{equation}
    S_{\rm gen}[\hat I(\epsilon)\cup R] = S_{\rm gen}[i(\epsilon)\cup R] +O(\epsilon^2)~.
\end{equation}

In $M$, the deformation profile $v^\mu\in N\partial \hat I$ can be decomposed as
\begin{align}
    v^\mu = -a k^{\mu} + b \ell^{\mu}~,
\end{align}
where $a$ and $b$ are positive definite functions. Hence
\begin{align}\label{eq-sign1}
    \left. \frac{d S_{\rm gen}[\hat I(\epsilon)\cup R]}{d\epsilon}\right|_{\epsilon=0} = \int dx \sqrt{h}~v^\mu \Theta_\mu[\hat{I}\cup R]~,
\end{align}
where $h$ refers to the intrinsic metric of $\partial \hat{I}$. We now choose $\tilde v\equiv 0$ outside a $\delta$-neighborhood of $q$ in $\partial \hat I$. For small enough $\delta$, $v^\mu \Theta_\mu[\hat{I}\cup R]<0$ in the entire $\delta$-neighborhood, by Eqs.~\eqref{knormal3} and \eqref{lnormal3} and continuity. Hence,
\begin{equation}
    \left. \frac{dS_{\rm gen}[\hat I(\epsilon)\cup R]}{d\epsilon}\right|_{\epsilon=0} = \left. \frac{dS_{\rm gen}[i(\epsilon)\cup R]}{d\epsilon}\right|_{\epsilon=0} <0~.
\end{equation}
Hence, $\hat I$ does not minimize the generalized entropy on $\Sigma$. This contradicts our assumption that $\hat I$ is a maximin region. Therefore, no such point $q\in \partial \hat I\cap\partial I_0$ can exist:
\begin{equation}
    \partial \hat{I} \cap \partial I_0=\varnothing~.
    \label{notini0}
\end{equation}

Suppose instead that $q\in \partial D(I_0)-\partial I_0$. In this case, minimality of $S_{\rm gen}(\hat I\cup R)$ on $\Sigma$, together with Eq.~\eqref{knormal3} or \eqref{lnormal3}, implies that a small deformation of $\Sigma$ into the interior of $D(I_0)$ near $q$ will produce a Cauchy surface of $I_0$ whose minimal-$S_{\rm gen}$ region (together with $R$) has greater generalized entropy than $\hat I\cup R$. But this is impossible if $\hat I$ was constructed by the maximin procedure. Again we will now aim to make this argument rigorous.
\begin{figure}[t]
\begin{center}
  \includegraphics[scale=1.9]{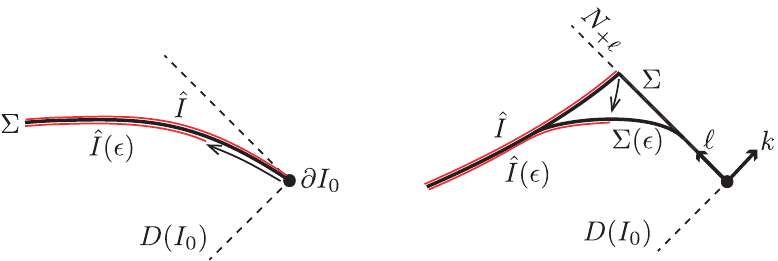} 
\end{center}
\caption{Maximin restricted to the domain of dependence (``wedge'') $D(I_0)$ returns a region $\hat I$ on a maximin slice $\Sigma$. If $I_0 \cup R$ is quantum normal, then $\partial\hat I$ cannot intersect $\partial D(I_0)$ (dashed). Left: if $\hat{I}\cap \partial I_0\neq\varnothing$, then $S_{\rm gen}(\hat{I}(\epsilon)\cup R) < S_{\rm gen}(\hat{I}\cup R)$, contradicting the min of maximin. Right: if $\hat{I}\cap \partial D(I_0)-\partial I_0$ then $S_{\rm gen}(\hat{I}(\epsilon)\cup R) > S_{\rm gen}(\hat{I}\cup R)$ on the deformed slice $\Sigma(\epsilon)$ violates the max of maximin.}
\label{fig:maximin}
\end{figure}

For definiteness, we assume that $q \in N_{+\ell}$. (If instead $q \in N_{-k}$, the time reverse of our argument applies.) In any open neighborhood $O(q)$, $\Sigma$ (and hence $\hat I$) must enter the interior, $O(q)\cap \Sigma \cap \text{int}[D(I_0)]\neq \varnothing$, or else Eq. \eqref{lnormal3} would violate the minimality of $S_{\rm gen}(\hat I\cup R)$ on $\Sigma$. Hence the inward-directed vector field $t^\mu$ orthogonal to $\partial \hat I$ and tangent to $\hat I$ is spacelike in an open neighborhood of $q$ on $\partial \hat I$. Moreover, $\Sigma$ must contain the null generator segment of $N_{+\ell}$ connecting $\partial I_0$ to $q$, or $\Sigma$ would fail to be achronal.

Let the achronal hypersurfaces $\Sigma(\epsilon)$ be a smooth one parameter deformation of $\Sigma$ such that $\Sigma(\epsilon)$ agrees with $\Sigma$ outside a $\delta$-neighborhood of $q$ (and everywhere for $\epsilon=0$). We also require that $\Sigma(\epsilon_2)$ is nowhere to the future of $\Sigma(\epsilon_1)$ if $\epsilon_1<\epsilon_2$. The stability assumption guarantees the existence of a smooth one-parameter family of regions $\hat I(\epsilon)$, each of which locally minimizes $S_{\rm gen}[\hat I(\epsilon)\cup R]$ on the corresponding $\Sigma(\epsilon)$. By the max step of maximin,
\begin{equation}
    \frac{dS_{\rm gen}[\hat I(\epsilon) \cup R]}{d\epsilon}\leq 0~.
    \label{stab}
\end{equation}
For small enough $\epsilon$, there exists an infinitesimal vector field $w^{\mu}$ in the normal bundle of $\partial \hat{I}$ in a neighborhood of $q$ such that 
\begin{equation}
        \partial \hat{I}(\epsilon) = \{\exp_{\partial \hat{I}}(p, \epsilon w^\mu(p)+O(\epsilon^2)): p \in \partial \hat{I} \}~.
\end{equation}
We have
\begin{equation}
    \frac{dS_{\rm gen}[\hat{I}(\epsilon)\cup R]}{d\epsilon} = \int dx \sqrt{h}~w^\mu \Theta_\mu{[\hat{I}\cup R]}~.
\end{equation}
Since $t^\mu$ is spacelike, it is linearly independent of $\ell^\mu$, so there exists a unique decomposition
\begin{equation}
     w^\mu = c t^\mu + d \,(-\ell^\mu)~.
\end{equation}
with $c$ and $d$ nonnegative functions on $\partial \hat{I}$ that vanish outside an open neighborhood of $q$. Hence
\begin{equation}
    \frac{S_{\rm gen}[\hat{I}\cup R]}{d\epsilon} = \int dx \sqrt{h}~[c t^\mu \Theta_\mu+ d\, (-\ell^\mu) \Theta_\mu]~.
\end{equation}
The first term is positive-definite by the minimality of $S_{\rm gen}[\hat I\cup R]$ on $\Sigma$; the second is positive by Eq.~\eqref{lnormal3}. Hence 
\begin{equation}
    \frac{S_{\rm gen}[\hat{I}\cup R]}{d\epsilon} > 0~,
\end{equation}
which contradicts Eq.~\eqref{stab}. Therefore,
\begin{equation}
   \partial \hat I\cap N_{+\ell} = \varnothing~,
\end{equation}
and a time-reversed argument implies
\begin{equation}
   \partial \hat I\cap N_{-k} = \varnothing~.
\end{equation}
Together with Eq.~\eqref{notini0} this establishes that
\begin{equation}
    \hat I\subset \text{int}[D(I_0)]~;
\end{equation}
hence $\hat I$ is locally quantum extremal.

\paragraph{Proof, part II} We will now discuss what happens if we relax the conditions \eqref{knormalstrict} and \eqref{lnormalstrict} to their non-strict versions \eqref{knormal} and \eqref{lnormal}.
We will argue that while in this case $\hat{I}$ might not be contained in $\text{int}[D(I_0)]$, an island candidate still exists, i.e., there exists $\hat{I} \subset D(I_0)$ such that $\hat{I} \cup R$ is quantum extremal and $S_{\rm gen}(\hat{I}\cup R)<S_{\rm gen}(R)$.


Let us start with the case where there exists at least a point $p \in \partial I_0$ where $k^\mu \Theta_\mu[I_0 \cup R;p]>0$ and a point $q$ ($p=q$ is allowed) where $\ell^\mu \Theta_\mu[I_0 \cup R;p]<0$. Then, $\partial\hat{I}$ cannot be a cross section of $N_{-k}$ ($N_{+\ell}$) because then by the QFC there would be a point $r$ in the cross section, along the same generator as $p$ ($q$), where $k^\mu \Theta_\mu[\hat{I} \cup R;r]>0$ ($\ell^\mu \Theta_\mu[\hat{I} \cup R;r]<0$). As discussed above, this contradicts maximin.

If $\partial \hat{I}$ only partially coincides with $\partial D(I_0)$ then there must exist a point $r$ in the boundary of the intersection set such that in any sufficiently small neighborhood of it $\partial \hat{I}$ and $N_{-k}$ ($N_{+\ell}$) do not coincide. In the $\hbar \to 0$ limit, Lemma B of \cite{C:2013uza} implies that there exists a point $s$ in a neighborhood of $r$ such that $k^\mu \theta_\mu[\hat{I};s]>0$ ($\ell^\mu \theta_\mu[\hat{I};s]<0$). Here, we will assume that the Lemma B of~\cite{C:2013uza} continues to hold when we replace the classical expansion with the quantum expansion.\footnote{Our assumption is motivated by the semiclassical generalization of many similar conditions on the classical expansion~\cite{Bousso:2015mna}. Note that if the von Neumann entropy term in the quantum expansion is $O(G\hbar)$ while the classical expansion is $O(1)$, Lemma B trivially generalizes to the version with quantum expansions.} We will then conclude that $k^\mu \Theta_\mu[\partial \hat{I} \cup R;s]>0$ ($\ell^\mu \Theta_\mu[\partial \hat{I} \cup R;s]<0$). However, $s \in \text{int}[D(I_0)]$, so a non-zero quantum expansion at $s$ is not allowed by maximin.

Next, we consider the case where $\ell^\mu \Theta_\mu[I_0 \cup R]=0$ (The case with $k$ and $\ell$ exchanged is similar by time-reflection symmetry). By the previous arguments, $\partial \hat{I}$ cannot intersect $N_{-k}$. Also, by the quantum version of Lemma B, $\partial \hat{I}$ cannot intersect $N_{+\ell}$ only in part. We will therefore focus on discussing the case where $\partial \hat{I}$ is a cross section of $N_{+\ell}$.

Associated with any cross section $L$ of $N_{+\ell}$, we can define a partial Cauchy slice $\Sigma_L$ of $D(I_0)$ which intersects $N_{+\ell}$ at $L$. Let $L_1$ be the latest cross section of $N_{+\ell}$ such that any other cross section $L$ in the past of $L_1$ satisfies $\ell^\mu \Theta_\mu[\Sigma_L \cup R]=0$.\footnote{If such $L_1$ does not exists, then $N_{+\ell}$ is a semi-infinite stationary null hypersurface which, at least classically, by the no-hair theorem needs to be a semi-infinite portion of the horizon of a Kerr-Newman black hole. And since $k^\mu \Theta_\mu[I_0 \cup R]\geq0$, $\partial I_0$ needs to lie fully in the past of the bifurcation surface which provides us with a classical extremal surface on $N_{+\ell}$. We then expect to find a quantum extremal region $\hat{I} \cup R$ such that $\partial \hat{I}$ is either on or near this classical extremal surface.} By the QFC, any cross section $L$ which is partly in the future of $L_1$ needs to contain at least a point $r$ at which $\ell^\mu \Theta_\mu[\Sigma_L \cup R;r]<0$. Hence, $\partial \hat{I}$ must be a cross section in the past of $L_1$. Furthermore, a maximin Cauchy slice $\Sigma$ corresponding to $\hat{I}$ cannot intersect the future of $L_1$ as it would violate the minimality of $S_{\rm gen}(\hat{I}\cup R)$. $\Sigma$ then has to leave $N_{+\ell}$ in the past of $L_1$ or on $L_1$. Let $L_2$ be the cross section at which $\Sigma$ leaves $N_{+\ell}$. Then, $k^\mu \Theta_\mu[\Sigma_{L_2} \cup R]\leq 0$ or else the min condition is violated. Since $k^\mu \Theta_\mu[I_0 \cup R]\geq 0$, we expect that there exists a cross section $L_3$ between $\partial I_0$ and $L_2$ such that $k^\mu \Theta_\mu[\Sigma_{L_3} \cup R]=0$. In the classical limit in particular, we expect that the results of~\cite{Andersson:2007gy} showing a similar existence on spacelike Cauchy surfaces can be applied here by taking a limit of spatial Cauchy surfaces that approach $N_{+\ell}$. Together with stationarity along $N_{+\ell}$, this would imply that $\Sigma_{L_3} \cup R$ is quantum extremal.

\subsection{Examples}
\label{examples}

We will now present some examples where the above sufficient conditions (i) and (ii) provide an efficient diagnostic for the existence of islands. We will require no detailed calculations of matter entropy and its derivatives, nor will we be forced to assume special symmetries, low dimensions of spacetime, or adopt other toy models. Our sufficient conditions establish the existence of an island and its key properties, at the cost of not exactly locating the island. 

In a final example, we show that neither of the two sufficient conditions can be eliminated.

\paragraph{Evaporating black hole after the Page time} For concreteness, we pick asymptotically flat boundary conditions, though our conclusion will not depend on this choice. We furthermore assume that the black hole mostly radiates massless particles, as will be the case if its initial mass is sufficiently large.

We consider an evaporating black hole formed in a pure state. At late times, the spacetime will be approximately spherically symmetric. Let $r=(A/4\pi)^{1/2}$ be the area radius of spheres. 
Near the horizon, the metric is well-approximated by
\begin{equation}\label{eq-invaidya}
ds^2 = -\left(1-\frac{r_s(v)}{r}\right) dv^2+2dv\, dr+r^2 d\Omega^2~,
\end{equation}
Due to evaporation, $r_s(v)$ decreases slowly with retarded time $v$: $dr_s/dv \sim -O(G\hbar/r_s^2)$. For $r \gg r_s$, the metric is well-approximated instead by the outgoing Vaidya metric \cite{Abdolrahimi:2016emo}, but this will not be important in our analysis.
\begin{figure}[t]
\begin{center}
  \includegraphics[scale=.85]{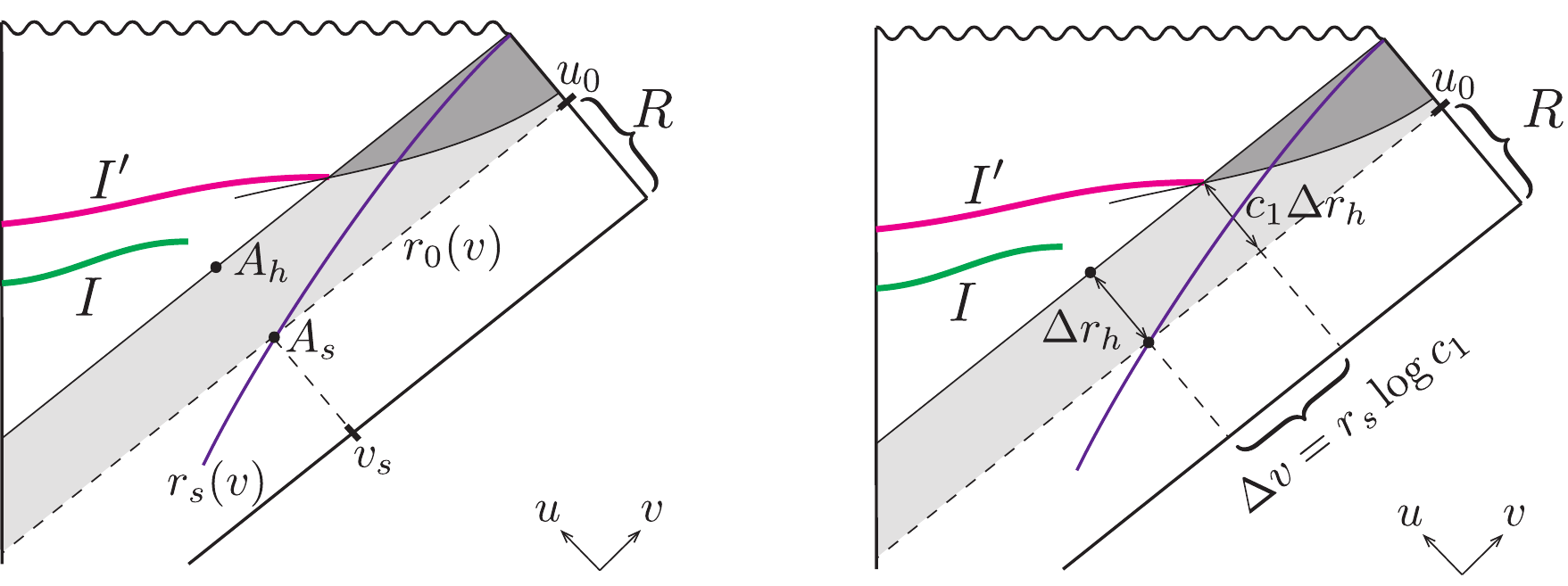} 
\end{center}
\caption{Evaporating black hole after the Page time. Hawking radiation has accumulated in $R$. As shown on the left, the boundary of the past of $R$, denoted by $r_0(v)$, intersects the stretched horizon (shown in purple) at the sphere $A_s$, which together with the $A_h$ sphere on the event horizons reside at retarded time $v_s$. We consider candidate regions $I'$ with boundary $\partial I'$ on a causal horizon spacelike to $R$ (grey regions). The generalized second law implies that $S_{\rm gen}(I'\cup R)$ increases under future outward deformations. For future inward deformations, quantum normalcy follows from the (trivial) classical normalcy in the dark grey subregion, which is chosen to keep quantum corrections to the expansion small. The $I'$ that minimizes $S_{\rm gen}(I'\cup R)$ subject to these restrictions is shown in pink. Its boundary is located a $\Delta v = r_s \log c_1$ to the future of $A_h$, as shown on the right. We show that it provides an extremely tight upper bound on the true entropy 
$S(\mathbf{R})=S_{\rm gen}(I\cup R)\leq S_{\rm gen}(I' \cup R) = A_h /4G\hbar + O(1)$.}
\label{fig:one-sided}
\end{figure}

Let $u$ be retarded time on $\mathscr{I}^+$, and let $R$ be the portion of $\mathscr{I}^+$ given by $u\leq u_0$; see Fig.~\ref{fig:one-sided}. $R$ is a reservoir that contains the Hawking radiation emitted until the time $u_0$. The boundary of the past of $R$ is given by $u=u_0$. We will be interested in the behavior of the metric only near the retarded time when this surface intersects the stretched horizon $r_s$, so it will be sufficient to set $r_s$ to that value and neglect its $v$-dependence from here on. Let $A_s=4\pi r_s^2$ be the area of the stretched horizon where it meets the past of $R$. Let $A_h=4\pi r_h^2$ be the area of the event horizon where it intersects the future of $A_s$; the areas satisfy
\begin{equation}
    A_s = A_h+O(G\hbar)~.
\end{equation}
We choose $u_0$ late enough so that $S(R)>A_h/4G\hbar+\log c_1$, where $\log c_1$ will be small in a sense made precise below. That is, $R$ extends to after the Page time, with a little room to spare. We seek an $I'$ that satisfies our sufficient conditions while placing a tight bound on the entropy $S(\mathbf{R})$.

Let $I'(r,v)$ be a Cauchy slice of the interior of the sphere $(r,v)$. Since $S_{\rm gen}(I' \cup R)$ is well-defined only for achronal $I' \cup R$, we require $I'(r,v) \subset M -J^-(R)$, or $u>u_0$. In the ingoing coordinates of Eq.~\eqref{eq-invaidya}, $u=u_0$ corresponds to a function $r_0(v)$, defined implicitly by $v=u_0+2 r_*(r_0)$, where $r_*(r)=r+r_h\log(\frac{r}{r_h}-1)$. Near the horizon, this satisfies
\begin{equation}
    \Delta r_0(v)\equiv r_0(v)-r_h=r_h \exp\left(\frac{v-u_0}{2r_h}-1\right)~.
    \label{rv}
\end{equation}
We thus require $r<r_0(v)$ for the boundary of $I'$.

Quantum normalcy of $I'\cup R$ requires that the generalized entropy grows along both of the null directions away from $I'$. Any future outward light-cone outside the horizon is a null surface of constant $u$ that reaches $\mathscr{I}^+$. Hence it is a causal horizon. The generalized second law of thermodynamics~\cite{Bekenstein:1972tm, Wall:2011hj} applies to all causal horizons. It implies that the future outward quantum expansion at $I'$ is positive if $\partial I'$ is outside the horizon. (In more general settings that are not exactly spherically symmetric, we can accomplish the same goal by choosing $\partial I'$ to be a cut of a causal horizon.) We thus require $r>r_h$.

The past outward {\em classical} expansion is trivially positive: $\partial_r A=8\pi r$. Quantum normalcy follows if the quantum correction, $4 G\hbar\, \partial_r S(I'\cup R)$, is negligible, i.e., if $\partial_r S(I'\cup R)\ll r/G\hbar$. Let $\Sigma$ be a global Cauchy slice containing $I'\cup R$ and define $I'_c = \Sigma-(I'\cup R)$. By purity of the global quantum state, $S(I'\cup R)=S(I'_c)$. 

Dimensional analysis dictates that the leading divergence in the renormalized entropy scales as 
\begin{equation}
    \partial_\rho S(I'_c)\sim O(1/\rho)~,
    \label{div}
\end{equation}
where $\rho=r_0-r$ and we may assume $\rho\ll r_0$. To see this, suppose first that the only available scales are $r$ and $\rho$. Terms stronger than Eq.~\eqref{div} would be of the form $\partial_r S(I'_c)\sim r^n/\rho^{n+1}$ with $n>0$ and positive coefficient. Such a term would imply that at fixed $\rho$, $dS/dr<0$, which is not physically sensible. In the presence of an additional mass scale $m\sim \hbar/\lambda$, an enhancement of Eq.~\eqref{div} would have to take the form $\partial_r S(I'_c)\sim \lambda^n/\rho^{n+1}$, $n>0$. Formally, this is an enhancement for $\rho\ll \lambda$, but physically, a mass scale cannot have any physical effect in this UV regime.

Hence, quantum normalcy is assured if we require
\begin{equation}
    r_0-r > c_1 \frac{G\hbar}{r}~,~~c_1\gg 1~. 
    \label{Deltar}
\end{equation}
To summarize, we may consider any $I'$ whose boundary is in the range
\begin{equation}
    r_h\leq r \leq r_0(v)+c_1\frac{G\hbar}{r}~,~~c_1\gg 1~. 
\end{equation}

We now minimize $S_{\rm gen}(I'\cup R)$ over this range. By purity, $S_{\rm gen}(I'\cup R)=4\pi r^2 +4G\hbar S(I'_c)$. Along any ingoing light-cone, the classical area decreases rapidly and $\Delta r$ only increases as we go to smaller $r$, so we are driven to the smallest $r$ in the search space, the event horizon. Scanning in the other null direction, along the event horizon, $S_{\rm gen}(I'_c)$ will decrease towards the past, by the GSL.

Hence we obtain the tightest upper bound on $S(\mathbf{R})$ by choosing $I'$ to be the interior of the event horizon, as early as is possible while maintaining Eq.~\eqref{Deltar}. With the boundary of $I'$ on the event horizon, $r-r_0=\Delta r_0\propto \exp(v/2r_h)$ by Eq.~\eqref{rv}. Moreover, at $A_h$, we have $r_0=r_s$ and hence $\Delta r_0\sim O(G\hbar/r_h)$. To grow this by the factor of $c_1$ demanded in Eq.~\eqref{Deltar}, we must choose $v=v_s+r_h\log c_1$, where $v_s$ is the $v$-coordinate of $A_s$ and $A_h$. 

To summarize, the optimal choice of $I'$ is
\begin{equation}
    (r,v)=(r_h, v_s+r_h\log c_1)~,~~c_1\gg 1~.
\end{equation}
The true entropy $S(\mathbf{R})$ is upper bounded by
\begin{equation}\label{eq-pagecurve}
    S_{\rm gen}(I'\cup R)= \frac{\pi r^2}{G\hbar}+ S(I'_c)=\frac{A_h}{4G\hbar}+ O(\log c_1)~.
\end{equation}
Note that the $O(G\hbar)$ area difference between the event horizon and the stretched horizon is negligible. The $O(\log c_1)$ term captures both the (negative) correction to the horizon area due to evaporation since $A_h$, and the (positive, and larger) correction due to the von Neumann entropy of $S(I'_c)$.

We stress that this upper bound is quite tight. The correct $S(\mathbf{R})$ is given by Eq.~\eqref{eq-pagecurve} with $O(\log c_1)$ replaced by $O(1)$. Recall that $c_1$ should be large enough to overcome any $O(1)$ coefficients that might enhance the von Neumann entropy in an exact calculation. But it is itself $O(1)$ in that sense, and $\log c_1$ is even smaller. In particular, we can always choose $\log c_1\ll \log \log(A_h/G\hbar)$ in the semiclassical limit.

We also emphasize that exact spherical symmetry is not crucial; our argument only relies on the scaling behavior of the relevant terms.

\paragraph{Recollapsing flat universe} Our next 
\begin{figure}[t]
\begin{center}
 \includegraphics[scale=.7]{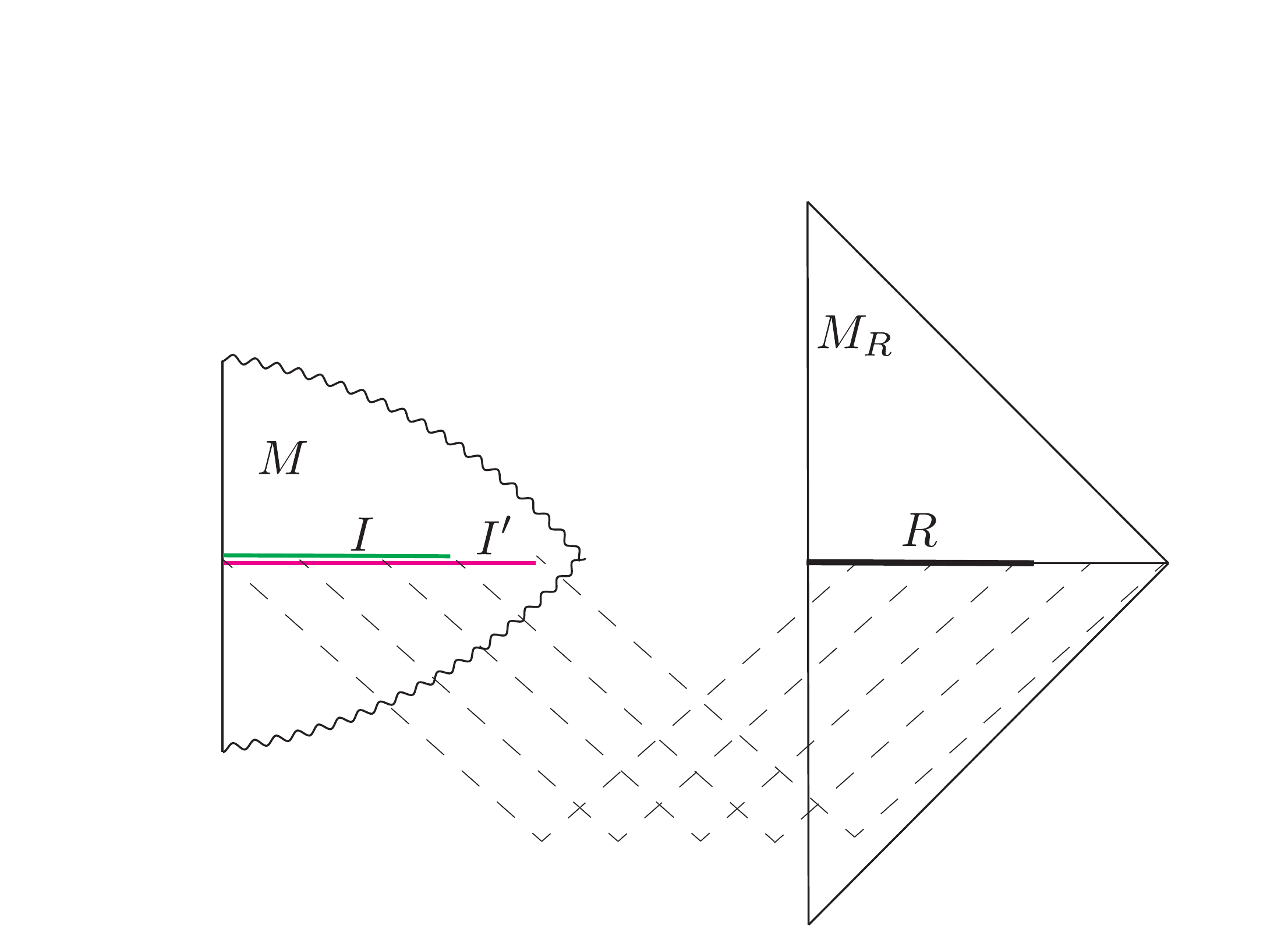}
\end{center}
\caption{Spatially flat radiation-dominated universe with negative cosmological constant, purified by a reference universe (thermal Minkowski space, right). If we choose a large enough reference region $R$ at $t_{\rm Mink}=0$, then the region $I'$ at the turnaround time $t=0$ satisfies our sufficient conditions. Therefore an island $I$ must exist.}
\label{fig:FRW}
\end{figure}
example was studied in detail in Ref.~\cite{Hartman:2020khs}.\footnote{Ref.~\cite{Hartman:2020khs} considered a different but related question to ours: given a region $I$ in the cosmology, can one find a region $R$ in the reference spacetime such that $I$ is an island with respect to $R$. By contrast, we specify a reference region $R$ and use our sufficient conditions to establish the existence of an island for it.} Consider a radiation-dominated, spatially flat Friedmann-Robertson-Walker (FRW) universe $M$ with cosmological constant $\Lambda$, purified by a thermal state on a Minkowski background $M_R$ without gravity. The metric of $M$ and $M_R$ is 
\begin{align}
    ds^2 &= -dt^2+a(t)^2 (dr^2+r^2\, d\Omega^2)~,\\
    ds_R^2 &= -dt_R^2+dr_R^2+r_R^2\, d\Omega_R^2~.
\end{align}
Without loss of generality, one can set the scale factor $a(0)=1$ at the turnaround time $t=0$, when $da/dt=0$ and hence $-\Lambda/8\pi G=\rho_{\rm rad}$. A thermofield double (TFD) state is constructed at $t=0$, $t_R=0$. 

A simple implication of the TFD state suffices for the purposes of our analysis. Consider two spatial regions, one in $M$ at $t=0$ and the other in $M_R$ at $t_R=0$. The von Neumann entropy of their union vanishes approximately, if they have the same spatial coordinates. If the regions are unequal, then the von Neumann entropy of their union will be given by the sum of the thermal entropy of the nonoverlap portions:
\begin{equation}
    S = s_{\rm rad} (\hat V+\hat V_R)~,
\end{equation}
where $\hat V$ and $\hat V_R$ are the volumes of the nonoverlap portions in $M$ and in $M_R$, and the entropy density is $s_{\rm rad}\sim \rho_{\rm rad}^{3/4}$. These statements receive corrections on scales below the thermal length scale, $\lambda\sim\rho_{\rm rad}^{-1/4}$.

Now choose $R\subset M_R$ to be a ball of radius $r_R$ at $t_R=0$, and $I'\subset M$ a ball of radius $r<r_R$ at $t=0$. By time symmetry around $t=t_R=0$, the region $I'\cup R$ will be quantum normal or anti-normal. We have
\begin{equation}
    S_{\rm gen}(I'\cup R)-S(R)= \frac{\pi r^2}{G\hbar}-\frac{4\pi s_{\rm rad}}{3} r^3~.
\end{equation}
To satisfy our second condition, this must be negative, so we require
\begin{equation}
    r > r_{\rm crit}\equiv \frac{3}{4\pi G s_{\rm rad}}
\end{equation}
This condition on $I'$ can be satisfied, and hence an island $I\subset M$ must exist, for a sufficiently large reference region, $r_R>r_{\rm crit}$.

Going beyond spherical symmetry, we can choose $R$ to be any convex reference region of arbitrary shape in $M_R$, and let $I'$ be the identical coordinate region in $M$. Then $I'\cup R$ will be normal or anti-normal by convexity and time symmetry. Moreover,
\begin{equation}
    S_{\rm gen}(I'\cup R)-S(R)= \frac{A[\partial I']}{4G\hbar}- \left(V_R+O(A[\partial R] \lambda)\right) s_{\rm rad}
\end{equation}
will be negative for any sufficiently large region of fixed shape. Any such references region must have an island $I$.

Note that $I$ will {\em not} be the identical coordinate region to $R$, because of the $O(A[\partial I'] \lambda)$ corrections to the von Neumann entropy. Moreover, in the nonspherical case, minimization of the area term will favor a more round shape for $I$ than for $R$.

\paragraph{Bag of gold} 
\begin{figure}[t]
\begin{center}
\includegraphics[scale=.5]{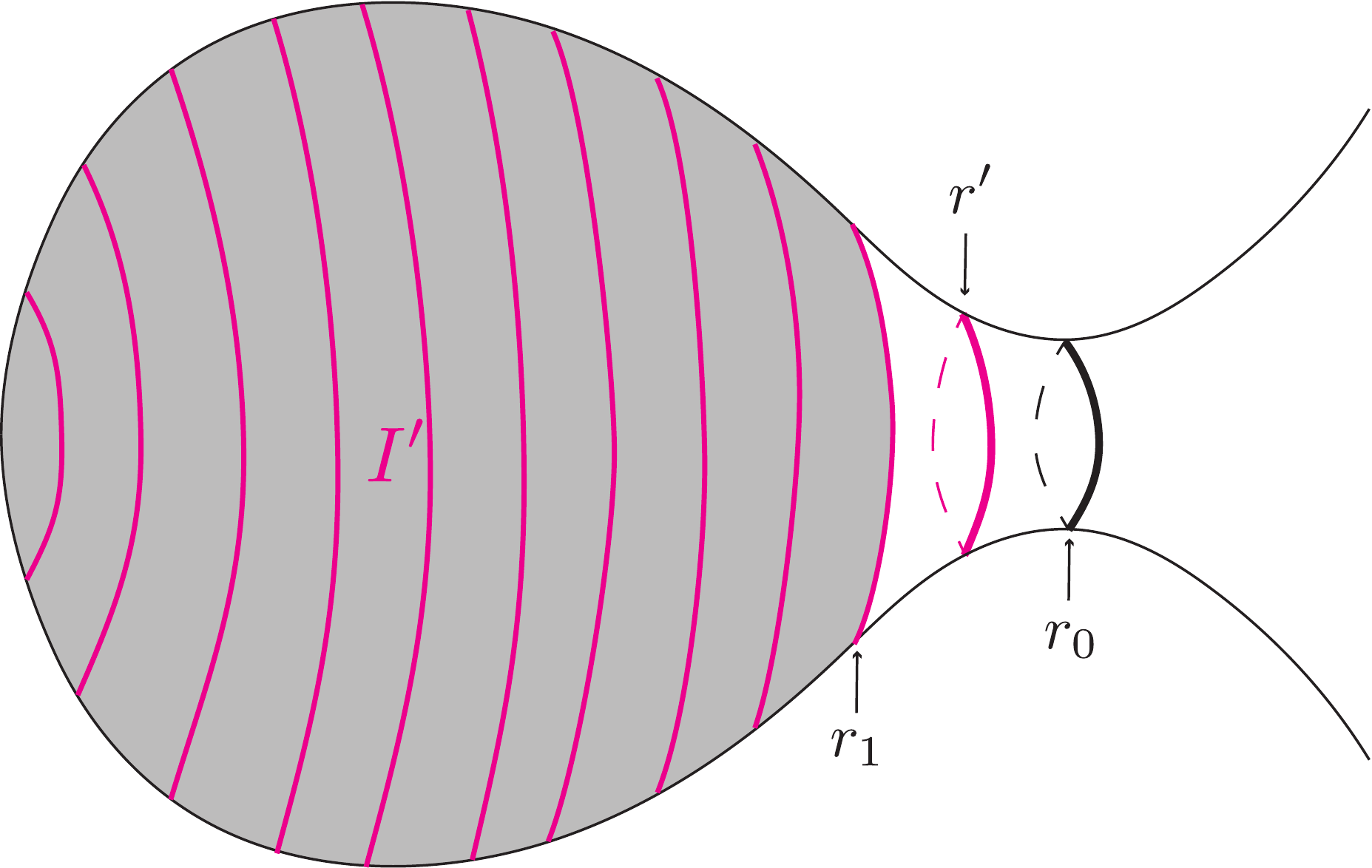}
\end{center}
\caption{A time-symmetric Cauchy slice of a bag-of-gold geometry. The bag has large entropy (grey) compared to the area of its throat, and it purifies the reference system $R$. Then it is easy to find a (classically and quantum) anti-normal region $I'$ (pink) such that $S_{\rm gen}(I'\cup R)<S(R)$. Hence, there must exist an island $I$. $I$ is expected to be approximately the interior of the classically minimal surface labelled $r_0$.}
\label{fig:BOG}
\end{figure}
Next, we discuss a time-symmetric slice of a ``bag-of-gold'' geometry~\cite{Marolf:2008tx}, shown in Fig. \ref{fig:BOG}. Its defining feature is the existence of an arbitrarily large volume of space behind a throat (a minimal area surface) of fixed area. To construct it, we glue the interior of a sphere of radius $r_1$ of a closed FRW universe, at the time of recollapse, $da/dt=0$, to the exterior of a sphere of the same size behind the bifurcation surface in a maximally extended Schwarzschild spacetime~\cite{Horowitz:1983vn}. The corresponding spatial metrics are:
\begin{align}
ds_{\rm in}^2 & = a (d\chi^2 + \sin^2 \chi d\Omega^2)~, ~~ & 0\leq \chi\leq \chi_1~;\\
ds_{\rm out}^2 & = 
\left(1-\frac{r_0}{r}\right)^{-1} dr^2 + r^2 d\Omega^2~,~~&r_0\leq r\leq r_1~,
\end{align}
where in the second line we omitted the portion of the Schwarzschild metric outside of the bifurcation surface as it will not be needed for the analysis below.

Let $r_0$ be the radius of the throat, and let $\chi$ be the angle at which the metrics are glued. The gravitational constraints imply $\chi_1>\pi/2$ and
\begin{align}
a \sin \chi_1 & = r_1 ~,\\
a \sin^3 \chi_1 & = r_0~.
\end{align}
The Friedmann equation implies that the energy density in the bag is
\begin{equation}
    \rho=\frac{1}{8\pi G a^2}~.
\end{equation}

Now suppose that the bag contains thermal photon radiation purified by a external reference system $R$. The entropy density in the bag is $s \sim \rho^{3/4}$, and hence 
\begin{equation}
S(R)\sim (G\hbar)^{-3/4}a^{3/2}~.
\end{equation}
Let $I'$ be the interior of some sphere $r'$ between the edge of the bag, $r_1$, and the throat, $r_0$; hence
\begin{equation}
 S_{\rm gen}(I' \cup R) \sim \frac{r'^2}{G\hbar}~.
\end{equation}

By time reversal symmetry, $I'\cup R$ is quantum normal or anti-normal. Moreover, we can achieve $S(R)>S_{\rm gen}(I' \cup R)$, by an arbitrarily large margin, by taking $a$ large while holding $r'$ and $r_0$ fixed. (This will only increase $r_1$.) Hence, our conditions are satisfied, and a nontrivial island $I \subset M$ must exist. 

Importantly, this construction is insensitive to the spherical symmetry that we assumed for simplicity. It is also insensitive to the addition of perturbative matter near the throat. Such modifications can affect the precise position of the island, which may be very hard to determine. But so long as they are small enough, our sufficient conditions will hold, and they guarantee the existence of an island.

\paragraph{Collapsing star (an example without islands)}

\begin{figure}[t]
\begin{center}
 \includegraphics[scale=.7]{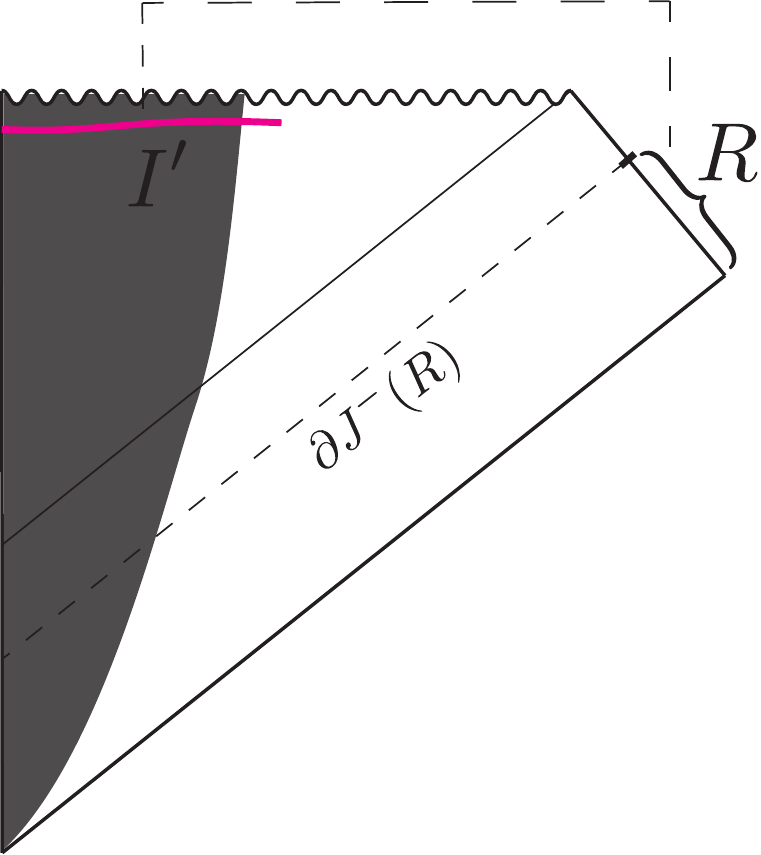}
\end{center}
\caption{Collapse of a spherical star that is maximally entangled with a distant reservoir $R$. $I'$ (pink) is the interior of a sphere surrounding the star at a time close to the singularity. Hence $\partial I'$ has a small area, and condition (i) can be satisfied by an arbitrarily large margin. But $I'$ is quantum trapped and so fails to satisfy condition (ii). Indeed, $R$ does not possess any island in this spacetime.}
\label{fig:noisland}
\end{figure}

To illustrate the importance of condition (ii), let us discuss a case for which condition (i), Eq.~\eqref{reduce}, is satisfied, but condition (ii), Eq.~\eqref{normal}, is violated. Consider an Oppenheimer–Snyder spacetime: a black hole formed by the collapse of a ``star,'' modeled as a spherical, homogeneous ball of dust. 

Suppose that the star is in a maximally mixed state with entropy $S_{\rm star}$ and let $R$ be an early portion of $\mathscr{I}^+$ which contains only a purification of the star (and no Hawking radiation), giving $S(R) = S_{\rm star}$. We choose $I'$ to be the interior of a sphere just outside of the star and very close to the singularity (see Fig.~\ref{fig:noisland}). Then $S_{\rm gen}(I'\cup R)\approx A(\partial I')/4G\hbar$. Picking $S_{\rm star}$ large with $A(\partial I')$ held fixed, we can arrange for
\begin{align}
    1 \ll S_{\rm gen}(I'\cup R) \ll S(R)~.
\end{align}
The first inequality ensures semiclassical control at $\partial I'$. The second states that condition (i) is satisfied (by an arbitrarily large margin). 

However, $\partial I'$ is a classically trapped surface, i.e. $\theta_k <0, \theta_\ell <0$. And since $\partial I'$ is not close to $\partial J^-(R)$, we expect quantum corrections to be small: $\Theta_{\ell} = \theta_{\ell}+ O(G\hbar)$. Condition (ii) is therefore violated.

Indeed, there are no islands associated to $R$ in this spacetime. To see this, note that there are no classically extremal spheres. As in the previous example, near $\partial J^-(R)$, quantum corrections to  $\theta_\ell$ can become large; but $\partial J^-(R)$ stays far from the horizon and so has large classical (and quantum) expansion everywhere.

This example shows that condition (ii) is essential. So is condition (i), of course. For example, suppose we chose $R$ to be a later portion of $\mathscr{I}^+$. As before, $R$ contains only the purification of the star, but no Hawking radiation. Since $\partial J^-(R)$ gets close to the horizon, where $\Theta_k$ can vanish, there will be a quantum extremal region $\tilde I$ with $\Theta_\ell$=0. However, this region fails to be an island because $S_{\rm gen}(\tilde{I} \cup R) > S(R)$.

\section{New Entropy Bound}
\label{neb}

In this section, we will show that in a globally pure state, the entropy of a reference system $R$ cannot exceed the generalized entropy of suitable asymptotic regions. We consider an external reference system in Sec.~\ref{neb1}, and we generalize to $R\subset M$ in Sec.~\ref{neb2}. We discuss examples in Sec.~\ref{nebex}.

\subsection{External Reference System}
\label{neb1}

Given an external reference system $R$, an island $I\subset M$ in a semiclassical spacetime $M$ is defined as a region that is quantum extremal and homologous to $R$ (i.e., $\partial I\subset M$), such that $S_{\rm gen}(I\cup R)$ is minimal among all such regions. In section~\ref{ext} we identified sufficient conditions for $I\neq \varnothing$.

We will now employ similar techniques to derive an entropy bound on the exact entropy of the reference system, $S(\mathbf{R})$, assuming that this is computed by the ``island formula'':
\begin{align}
    S(\mathbf{R}) = S_{\rm gen}(I\cup R)~.
    \label{islandrule}
\end{align}
Following Ref.~\cite{Almheiri:2019hni}, we denote $\mathbf{R}$ in boldface when referring to the exact (nonperturbatively computed) state of the region. We write $R$ when referring to the semiclassically computed state. For simplicity, we will assume that the global quantum state is pure,
\begin{equation}
    S(R\cup M)=0~, \label{pure}
\end{equation}
though generalizations can easily be considered. 

Let $I'_c \subset{M}$ be any partial Cauchy slice of $M$ that is quantum normal or anti-normal: 
\begin{equation}\label{eq-boundcondition1}
    \begin{cases}
    k^{\mu} \Theta_{\mu}[I'_c] \geq 0~,~~ \ell^{\mu} \Theta_{\mu}[I'_c] \leq 0~;\\
    \text{or}\\
    k^{\mu} \Theta_{\mu}[I'_c] \leq 0~,~~ \ell^{\mu} \Theta_{\mu}[I'_c] \geq 0~,
    \end{cases}
\end{equation}
where $k$ and $\ell$ are the future-directed null vector fields orthogonal to $\partial I'_c$. We also require that $I'_c$ is ``asymptotic,'' though only in the weak sense that in the conformally compactified spacetime,
\begin{equation}
    \partial \Sigma \subset\partial I'_c~,
    \label{homic}
\end{equation} 
where $\Sigma$ is a Cauchy slice of $M$. That is, $I'_c$ must contain the asymptotic region of $M$, but it may extend deep into the interior of $M$. A simple example of a region $I'_c$ that satisfies Eqs.~\eqref{eq-boundcondition1} and \eqref{homic} is the exterior of a sufficiently large approximately round sphere. 

Let $I'$ be the complement of $I'_c$ on some global Cauchy slice of $M$. By Eqs.~\eqref{pure} and \eqref{eq-boundcondition1}, $I'\cup R$ will be anti-normal or normal. By Eq.~\eqref{homic}, $I'$ is homologous to $R$. Our notation reflects the fact that $I'$ shares these properties with the region denoted $I'$ in Sec.~\ref{ext}. 

However, here we do {\em not} assume the inequality $S_{\rm gen}(I'\cup R)<S(R)$, and hence we will {\em not} be guaranteed the existence of an island $I\neq \varnothing$. This does not affect the maximin analysis performed in Sec.~\ref{ext}: (anti-)normalcy of $I'\cup R$ implies that the true island $I$ satisfies
\begin{equation}
    S_{\rm gen}(I\cup R)\leq S_{\rm gen}(I'\cup R)~,
\end{equation}
regardless of whether $I$ is the empty set or not. Using Eqs.~\eqref{islandrule} and \eqref{pure}, we thus find the entropy bound
\begin{equation}
    S(\mathbf{R}) \leq S_{\rm gen}(I'_c)~.
    \label{boundext}
\end{equation}

\subsection{Distant Reference System}
\label{neb2}

The bound \eqref{boundext} generalizes to the case where $R \subset M$, subject to appropriate modifications. (It is easy to generalize further to the case where $R$ is partly internal to $M$ and partly an external system.) We shall assume that gravity is negligible in $R$, so that the notion of an exact state of $\mathbf{R}$ can be made precise. The island rule can then be adapted to compute the generalized entropy of $R$:
\begin{align}
    S_{\rm gen}(\mathbf{R}) = S_{\rm gen}(I\cup R)~,
    \label{islandruleint}
\end{align}
where $I$ is an island (possibly the empty set), as described above. The relevant homology rule is $I\subset \mbox{int}[M-\overline{J(R)}]$, where $J$ denotes the union of the causal past and future.

\begin{figure}[t]
\begin{center}
 \includegraphics[scale=.5]{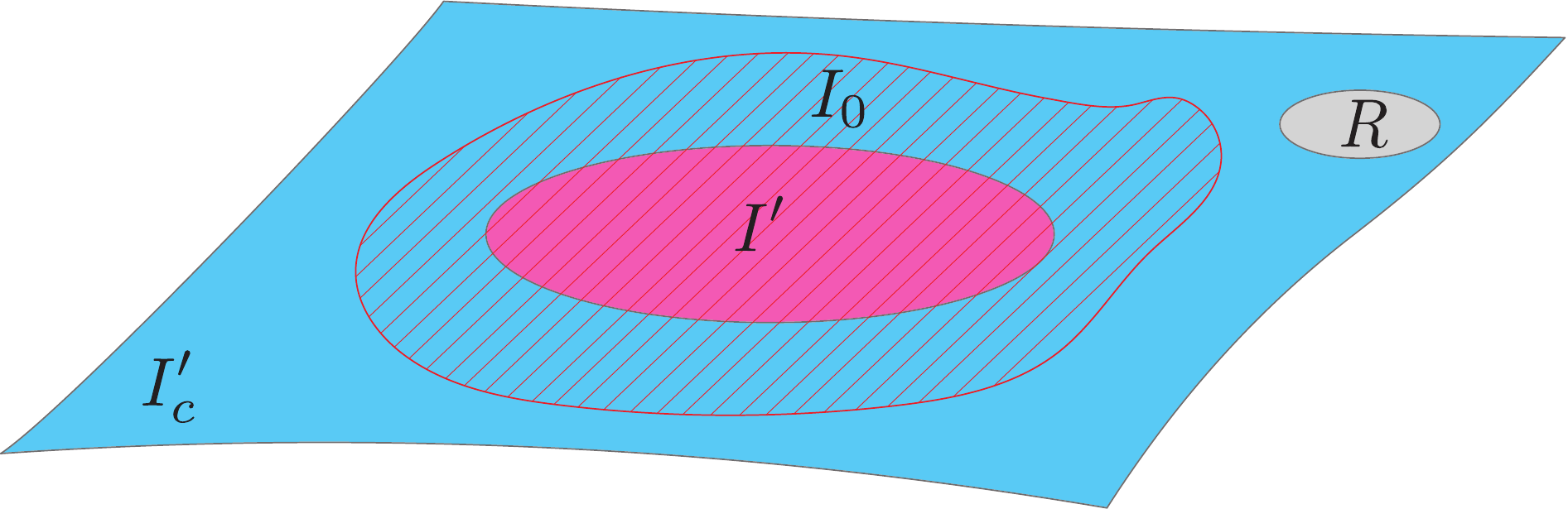} 
\end{center}
\caption{The entropy $S(\mathbf{R})$ of an external or distant reference system $R$ must be less than the generalized entropy of any region $I'_c$ that is normal or anti-normal (blue).}
\label{fig:Iprimecinternal}
\end{figure}

We again assume global purity, $S(M)=0$. To obtain a bound on $S_{\rm gen}(\mathbf{R})$, we consider a spatial region $I'_c$ that satisfies the following conditions (see Fig.~\ref{fig:Iprimecinternal}):
\begin{itemize}
    \item For $I'$ to be of the correct homology type, without directly referring to $I'$,\footnote{Our goal is to formulate a bound in terms of quantities that are accessible to an asymptotic observer.} we require that $I'_c$ is adjacent to $R$ in $M$; and in the conformally compactified spacetime $\tilde M$, $I'_c$ contains any conformal boundary portions not covered by $R$:
    \begin{equation}
    \partial I'_c\supset \partial(\tilde\Sigma - \bar R)~,
    \end{equation}
    where $\tilde \Sigma\supset R$ is a Cauchy slice of $\tilde M$, and $\bar R$ denotes the closure of $R$ in $\tilde M$.
    \item $I'_c$ is quantum normal or anti-normal under shape deformations of its inner boundary in $M$, i.e., at $(\partial I'_c-\partial R)\cap M$~.
    \item $I'_c$ contains a region $I_{0,c}$ that is quantum anti-normal at $\partial I_{0,c}-\partial R$. (Normal is not allowed in this criterion.)
\end{itemize}

Global purity implies that $I'\cup R$ will be quantum normal or anti-normal at $\partial I'$. It also guarantees quantum normalcy of $I_0 \cup R$, where $I_0\equiv \Sigma-I_{0,c}-R$ and $\Sigma$ is a Cauchy surface that contains $I_{0,c}$ and $R$. By Sec.~\ref{dist}, the maximin procedure restricted to the wedge $D(I_0)$ will return a region $\hat I\subset \mbox{int}[D(I_0)]$ that satisfies the homology rule and has stationary $S_{\rm gen}(\hat I\cup R)$ under shape deformations at $\partial \hat I$.

Note that $\hat I$ may be empty; and $S_{\rm gen}(\hat I\cup R)$ need not be globally minimal, since the true island $I$ may not be contained in $D(I_0)$. However, we have
\begin{equation}
    S_{\rm gen}(I\cup R)\leq S_{\rm gen}(\hat I\cup R)~,
\end{equation}
The quantum (anti-)normalcy of $I'$ implies 
\begin{equation}
    S_{\rm gen}(\hat I\cup R)\leq S_{\rm gen}(I'\cup R)~.
\end{equation}
Using Eqs.~\eqref{islandruleint} and global purity, we thus find the entropy bound
\begin{equation}
    S_{\rm gen}(\mathbf{R}) \leq S_{\rm gen}(I'_c)~.
    \label{boundint}
\end{equation}

Recall that $\partial I'_c\supset \partial R$, so in any situation where the generalized entropy can be separated into a regularized entropy and Bekenstein-Hawking term, the area terms associated with $\partial R$ will cancel, and Eq.~\eqref{boundint} reduces to Eq.~\eqref{boundext}.

\subsection{Examples and Discussion}
\label{nebex}

The bound \eqref{boundint}, and its external $R$ version \eqref{boundext}, are powerful and versatile. They require knowledge only of $R$ and $I'_c$, but not of the rest of the spacetime $M$. The only nontrivial condition, Eq.~\eqref{eq-boundcondition1}, can be easily verified. Often the quantum expansion is dominated by the classical expansion, so that it is easy to check whether $I'_c$ is quantum (anti-)normal; yet the bound remains nontrivial.
\begin{figure}[t]
\begin{center}
  \includegraphics[scale=1]{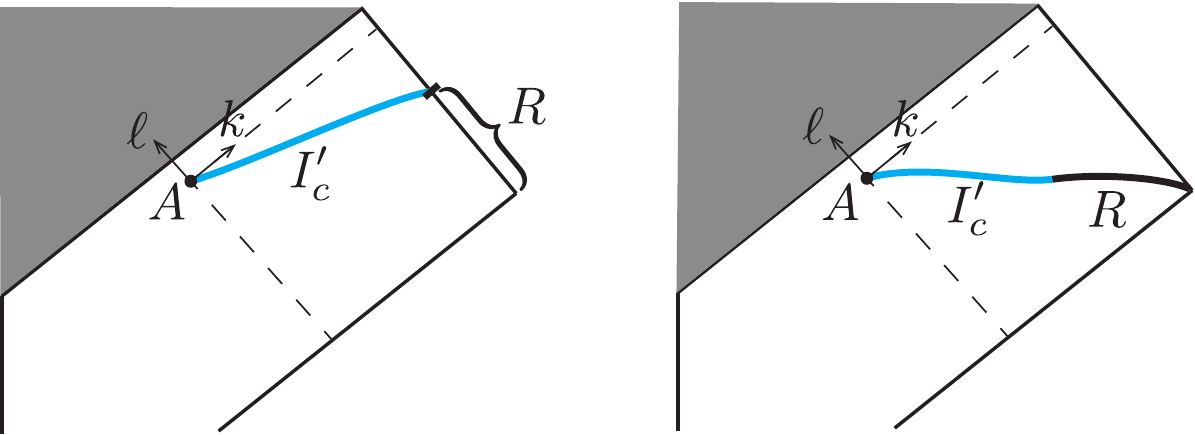} 
\end{center}
\caption{{\em Not} a Schwarzschild black hole. This could be a highly dynamical spacetime far from a stationary black hole solution. $I'_c$ is is quantum anti-normal, so $S(\mathbf{R})\leq S_{\rm gen}(I'_c)$. In the left example, $R$ is on the conformal boundary. Quantum anti-normalcy follows from the GSL (future null congruence), and from the classical area theorem (past null congruence) if quantum corrections are small. In the right example, $R$ is inside the spacetime. Quantum anti-normalcy follows by the area theorem and smallness of corrections if $\partial I'_c$ stays far enough from the horizon.}
\label{fig:one-sided-entropybound}
\end{figure}

For example, by the generalized second law, causal wedges of a boundary region must be quantum anti-normal.\footnote{Note that the causal wedge is not in general a domain of dependence. The region that is always quantum normal is the maximal Cauchy evolution of the causal wedge.} This follows both for asymptotically anti-de Sitter and flat spacetimes. When $R$ is disjoint from the conformal completion of $M$, this ensures the quantum anti-normalcy of $I'_c$. 

The requirement that $R$ and $I'_c$ be spacelike separated prevents the application of the GSL when $R$ is part of the conformal completion of $M$. For instance, suppose that $R$ is a subset of $\mathscr{I}^+$. The GSL can still be applied to any future causal horizon associated to $\mathscr{I}^+$ regions including $R$, guaranteeing $k^\mu \Theta_{\mu}\geq 0$. However, all past horizons will intersect the past of $R$, blocking the application of the GSL to $I'_c$ along them. To establish that $\ell^\mu\Theta_{\mu}\leq 0$, we can use the classical area law on past horizons, so long as quantum corrections to $\ell^\mu\theta_{\mu}$ is negligible (see Fig.~\ref{fig:one-sided-entropybound}, left). 

More generally, when $R$ is a subset of $M$, the classical area law ensures condition \eqref{eq-boundcondition1} if the quantum corrections to both expansions are suppressed (see right Fig.~\ref{fig:one-sided-entropybound}). This causal wedge method for finding $I'_c$ suggests a nice physical interpretation of $I'_c$ as a region that can be explored geometrically by asymptotic observers.

The bound thus tells us that $S(\mathbf{R})$ cannot be greater than the generalized entropy of any causal wedge region (subject to quantum effects on the expansion remaining negligible). If $I'_c$ is a whole Cauchy surface $\Sigma$ of $M$, this reduces to the trivial statement that $S(\mathbf{R})\leq S(\Sigma)$. (In this case, by purity, equality must hold.) But if $I'_c$ has a boundary in $M$, the bound is nontrivial. Indeed, a quantum anti-normal causal wedge can reach very close ($O[(G\hbar)^{1/2}]$ distance) to a black hole horizon. For a black hole after the Page time, this means that bound becomes nearly saturated. The bound then implies the nontrivial statement of unitarity. 

The bound simplifies if $I'_c$ contains little matter entropy, so that
\begin{equation}
     S_{\rm gen}(I'_c) \approx \frac{A}{4G\hbar}~,
\end{equation}
In this case,
\begin{equation}
     S(\mathbf{R}) \lesssim \frac{A}{4G\hbar}~.
\end{equation}
The entropy deliverable to an asymptotic observer by a spacetime causally explored to an inner boundary of area $A$ cannot be greater than $A/4G\hbar$.

\section*{Acknowledgements}
ASM would like to thank Raghu Mahajan for discussions. This work was supported in part by the Berkeley Center for Theoretical Physics; by the Department of Energy, Office of Science, Office of High Energy Physics under QuantISED Award DE-SC0019380 and under contract DE-AC02-05CH11231; and by the National Science Foundation under Award Numbers 1820912 (RB) and 2014215 (ASM).

\bibliographystyle{JHEP}
\bibliography{main}

\providecommand{\href}[2]{#2}\begingroup\raggedright\begin{thebibliography}{10}

\bibitem{Engelhardt:2014gca}
N.~Engelhardt and A.~C. Wall, \emph{{Quantum Extremal Surfaces: Holographic
  Entanglement Entropy beyond the Classical Regime}},
  \href{https://doi.org/10.1007/JHEP01(2015)073}{\emph{JHEP} {\bfseries 01}
  (2015) 073} [\href{https://arxiv.org/abs/1408.3203}{{\ttfamily 1408.3203}}].

\bibitem{Hubeny:2007xt}
V.~E. Hubeny, M.~Rangamani and T.~Takayanagi, \emph{{A Covariant holographic
  entanglement entropy proposal}},
  \href{https://doi.org/10.1088/1126-6708/2007/07/062}{\emph{JHEP} {\bfseries
  07} (2007) 062} [\href{https://arxiv.org/abs/0705.0016}{{\ttfamily
  0705.0016}}].

\bibitem{Ryu:2006bv}
S.~Ryu and T.~Takayanagi, \emph{{Holographic derivation of entanglement entropy
  from AdS/CFT}},
  \href{https://doi.org/10.1103/PhysRevLett.96.181602}{\emph{Phys. Rev. Lett.}
  {\bfseries 96} (2006) 181602}
  [\href{https://arxiv.org/abs/hep-th/0603001}{{\ttfamily hep-th/0603001}}].

\bibitem{Lewkowycz:2013nqa}
A.~Lewkowycz and J.~Maldacena, \emph{{Generalized gravitational entropy}},
  \href{https://doi.org/10.1007/JHEP08(2013)090}{\emph{JHEP} {\bfseries 08}
  (2013) 090} [\href{https://arxiv.org/abs/1304.4926}{{\ttfamily 1304.4926}}].

\bibitem{Faulkner:2013ana}
T.~Faulkner, A.~Lewkowycz and J.~Maldacena, \emph{{Quantum corrections to
  holographic entanglement entropy}},
  \href{https://doi.org/10.1007/JHEP11(2013)074}{\emph{JHEP} {\bfseries 11}
  (2013) 074} [\href{https://arxiv.org/abs/1307.2892}{{\ttfamily 1307.2892}}].

\bibitem{Dong:2017xht}
X.~Dong and A.~Lewkowycz, \emph{{Entropy, Extremality, Euclidean Variations,
  and the Equations of Motion}},
  \href{https://doi.org/10.1007/JHEP01(2018)081}{\emph{JHEP} {\bfseries 01}
  (2018) 081} [\href{https://arxiv.org/abs/1705.08453}{{\ttfamily
  1705.08453}}].

\bibitem{Penington:2019kki}
G.~Penington, S.~H. Shenker, D.~Stanford and Z.~Yang, \emph{{Replica wormholes
  and the black hole interior}},
  \href{https://arxiv.org/abs/1911.11977}{{\ttfamily 1911.11977}}.

\bibitem{Almheiri:2019qdq}
A.~Almheiri, T.~Hartman, J.~Maldacena, E.~Shaghoulian and A.~Tajdini,
  \emph{{Replica Wormholes and the Entropy of Hawking Radiation}},
  \href{https://doi.org/10.1007/JHEP05(2020)013}{\emph{JHEP} {\bfseries 05}
  (2020) 013} [\href{https://arxiv.org/abs/1911.12333}{{\ttfamily
  1911.12333}}].

\bibitem{Page:1993wv}
D.~N. Page, \emph{{Information in black hole radiation}},
  \href{https://doi.org/10.1103/PhysRevLett.71.3743}{\emph{Phys. Rev. Lett.}
  {\bfseries 71} (1993) 3743}
  [\href{https://arxiv.org/abs/hep-th/9306083}{{\ttfamily hep-th/9306083}}].

\bibitem{Penington:2019npb}
G.~Penington, \emph{{Entanglement Wedge Reconstruction and the Information
  Paradox}}, \href{https://doi.org/10.1007/JHEP09(2020)002}{\emph{JHEP}
  {\bfseries 09} (2020) 002}
  [\href{https://arxiv.org/abs/1905.08255}{{\ttfamily 1905.08255}}].

\bibitem{Almheiri:2019psf}
A.~Almheiri, N.~Engelhardt, D.~Marolf and H.~Maxfield, \emph{{The entropy of
  bulk quantum fields and the entanglement wedge of an evaporating black
  hole}}, \href{https://doi.org/10.1007/JHEP12(2019)063}{\emph{JHEP} {\bfseries
  12} (2019) 063} [\href{https://arxiv.org/abs/1905.08762}{{\ttfamily
  1905.08762}}].

\bibitem{Hawking:1974sw}
S.~W. Hawking, \emph{{Particle Creation by Black Holes}},
  \href{https://doi.org/10.1007/BF02345020}{\emph{Commun. Math. Phys.}
  {\bfseries 43} (1975) 199}.

\bibitem{Hawking:1976ra}
S.~W. Hawking, \emph{{Breakdown of Predictability in Gravitational Collapse}},
  \href{https://doi.org/10.1103/PhysRevD.14.2460}{\emph{Phys. Rev. D}
  {\bfseries 14} (1976) 2460}.

\bibitem{Bousso:2015mna}
R.~Bousso, Z.~Fisher, S.~Leichenauer and A.~C. Wall, \emph{{Quantum focusing
  conjecture}}, \href{https://doi.org/10.1103/PhysRevD.93.064044}{\emph{Phys.
  Rev.} {\bfseries D93} (2016) 064044}
  [\href{https://arxiv.org/abs/1506.02669}{{\ttfamily 1506.02669}}].

\bibitem{Almheiri:2019yqk}
A.~Almheiri, R.~Mahajan and J.~Maldacena, \emph{{Islands outside the horizon}},
   \href{https://arxiv.org/abs/1910.11077}{{\ttfamily 1910.11077}}.

\bibitem{Bousso:2019ykv}
R.~Bousso and M.~Toma\v{s}evi\'c, \emph{{Unitarity From a Smooth Horizon?}},
  \href{https://doi.org/10.1103/PhysRevD.102.106019}{\emph{Phys. Rev. D}
  {\bfseries 102} (2020) 106019}
  [\href{https://arxiv.org/abs/1911.06305}{{\ttfamily 1911.06305}}].

\bibitem{Bousso:2020kmy}
R.~Bousso and E.~Wildenhain, \emph{{Gravity/ensemble duality}},
  \href{https://doi.org/10.1103/PhysRevD.102.066005}{\emph{Phys. Rev. D}
  {\bfseries 102} (2020) 066005}
  [\href{https://arxiv.org/abs/2006.16289}{{\ttfamily 2006.16289}}].

\bibitem{Dong:2020uxp}
X.~Dong, X.-L. Qi, Z.~Shangnan and Z.~Yang, \emph{{Effective entropy of quantum
  fields coupled with gravity}},
  \href{https://doi.org/10.1007/JHEP10(2020)052}{\emph{JHEP} {\bfseries 10}
  (2020) 052} [\href{https://arxiv.org/abs/2007.02987}{{\ttfamily
  2007.02987}}].

\bibitem{Gautason:2020tmk}
F.~F. Gautason, L.~Schneiderbauer, W.~Sybesma and L.~Thorlacius, \emph{{Page
  Curve for an Evaporating Black Hole}},
  \href{https://doi.org/10.1007/JHEP05(2020)091}{\emph{JHEP} {\bfseries 05}
  (2020) 091} [\href{https://arxiv.org/abs/2004.00598}{{\ttfamily
  2004.00598}}].

\bibitem{Hartman:2020swn}
T.~Hartman, E.~Shaghoulian and A.~Strominger, \emph{{Islands in Asymptotically
  Flat 2D Gravity}}, \href{https://doi.org/10.1007/JHEP07(2020)022}{\emph{JHEP}
  {\bfseries 07} (2020) 022}
  [\href{https://arxiv.org/abs/2004.13857}{{\ttfamily 2004.13857}}].

\bibitem{Almheiri:2019hni}
A.~Almheiri, R.~Mahajan, J.~Maldacena and Y.~Zhao, \emph{{The Page curve of
  Hawking radiation from semiclassical geometry}},
  \href{https://doi.org/10.1007/JHEP03(2020)149}{\emph{JHEP} {\bfseries 03}
  (2020) 149} [\href{https://arxiv.org/abs/1908.10996}{{\ttfamily
  1908.10996}}].

\bibitem{Almheiri:2019psy}
A.~Almheiri, R.~Mahajan and J.~E. Santos, \emph{{Entanglement islands in higher
  dimensions}},
  \href{https://doi.org/10.21468/SciPostPhys.9.1.001}{\emph{SciPost Phys.}
  {\bfseries 9} (2020) 001} [\href{https://arxiv.org/abs/1911.09666}{{\ttfamily
  1911.09666}}].

\bibitem{Chen:2020tes}
Y.~Chen, V.~Gorbenko and J.~Maldacena, \emph{{Bra-ket wormholes in
  gravitationally prepared states}},
  \href{https://arxiv.org/abs/2007.16091}{{\ttfamily 2007.16091}}.

\bibitem{Hartman:2020khs}
T.~Hartman, Y.~Jiang and E.~Shaghoulian, \emph{{Islands in cosmology}},
  \href{https://doi.org/10.1007/JHEP11(2020)111}{\emph{JHEP} {\bfseries 11}
  (2020) 111} [\href{https://arxiv.org/abs/2008.01022}{{\ttfamily
  2008.01022}}].

\bibitem{Wall:2012uf}
A.~C. Wall, \emph{{Maximin Surfaces, and the Strong Subadditivity of the
  Covariant Holographic Entanglement Entropy}},
  \href{https://doi.org/10.1088/0264-9381/31/22/225007}{\emph{Class. Quant.
  Grav.} {\bfseries 31} (2014) 225007}
  [\href{https://arxiv.org/abs/1211.3494}{{\ttfamily 1211.3494}}].

\bibitem{Akers:2019lzs}
C.~Akers, N.~Engelhardt, G.~Penington and M.~Usatyuk, \emph{{Quantum Maximin
  Surfaces}}, \href{https://doi.org/10.1007/JHEP08(2020)140}{\emph{JHEP}
  {\bfseries 08} (2020) 140}
  [\href{https://arxiv.org/abs/1912.02799}{{\ttfamily 1912.02799}}].

\bibitem{Marolf:2019bgj}
D.~Marolf, A.~C. Wall and Z.~Wang, \emph{{Restricted Maximin surfaces and HRT
  in generic black hole spacetimes}},
  \href{https://doi.org/10.1007/JHEP05(2019)127}{\emph{JHEP} {\bfseries 05}
  (2019) 127} [\href{https://arxiv.org/abs/1901.03879}{{\ttfamily
  1901.03879}}].

\bibitem{Brown:2019rox}
A.~R. Brown, H.~Gharibyan, G.~Penington and L.~Susskind, \emph{{The
  Python\textquoteright{}s Lunch: geometric obstructions to decoding Hawking
  radiation}}, \href{https://doi.org/10.1007/JHEP08(2020)121}{\emph{JHEP}
  {\bfseries 08} (2020) 121}
  [\href{https://arxiv.org/abs/1912.00228}{{\ttfamily 1912.00228}}].

\bibitem{Akers:2017nrr}
C.~Akers, R.~Bousso, I.~F. Halpern and G.~N. Remmen, \emph{{Boundary of the
  future of a surface}},
  \href{https://doi.org/10.1103/PhysRevD.97.024018}{\emph{Phys. Rev. D}
  {\bfseries 97} (2018) 024018}
  [\href{https://arxiv.org/abs/1711.06689}{{\ttfamily 1711.06689}}].

\bibitem{C:2013uza}
A.~C. Wall, \emph{{The Generalized Second Law implies a Quantum Singularity
  Theorem}}, \href{https://doi.org/10.1088/0264-9381/30/19/199501}{\emph{Class.
  Quant. Grav.} {\bfseries 30} (2013) 165003}
  [\href{https://arxiv.org/abs/1010.5513}{{\ttfamily 1010.5513}}].

\bibitem{Andersson:2007gy}
L.~Andersson and J.~Metzger, \emph{{The Area of horizons and the trapped
  region}}, \href{https://doi.org/10.1007/s00220-008-0723-y}{\emph{Commun.
  Math. Phys.} {\bfseries 290} (2009) 941}
  [\href{https://arxiv.org/abs/0708.4252}{{\ttfamily 0708.4252}}].

\bibitem{Abdolrahimi:2016emo}
S.~Abdolrahimi, D.~N. Page and C.~Tzounis, \emph{{Ingoing Eddington-Finkelstein
  Metric of an Evaporating Black Hole}},
  \href{https://doi.org/10.1103/PhysRevD.100.124038}{\emph{Phys. Rev. D}
  {\bfseries 100} (2019) 124038}
  [\href{https://arxiv.org/abs/1607.05280}{{\ttfamily 1607.05280}}].

\bibitem{Bekenstein:1972tm}
J.~D. Bekenstein, \emph{{Black holes and the second law}},
  \href{https://doi.org/10.1007/BF02757029}{\emph{Lett. Nuovo Cim.} {\bfseries
  4} (1972) 737}.

\bibitem{Wall:2011hj}
A.~C. Wall, \emph{{A proof of the generalized second law for rapidly changing
  fields and arbitrary horizon slices}},
  \href{https://doi.org/10.1103/PhysRevD.85.104049}{\emph{Phys. Rev. D}
  {\bfseries 85} (2012) 104049}
  [\href{https://arxiv.org/abs/1105.3445}{{\ttfamily 1105.3445}}].

\bibitem{Marolf:2008tx}
D.~Marolf, \emph{{Black Holes, AdS, and CFTs}},
  \href{https://doi.org/10.1007/s10714-008-0749-7}{\emph{Gen. Rel. Grav.}
  {\bfseries 41} (2009) 903} [\href{https://arxiv.org/abs/0810.4886}{{\ttfamily
  0810.4886}}].

\bibitem{Horowitz:1983vn}
G.~T. Horowitz, \emph{{The positive energy theorem and its extensions}},  in
  \emph{{Conference on Asymptotic Behavior of Mass and Spacetime Geometry}},
  p.~0001, 10, 1983.

\end{thebibliography}\endgroup



\providecommand{\href}[2]{#2}\begingroup\raggedright\endgroup

\end{document}